\magnification1200

\vskip .5cm
\centerline {\bf Poisson equations, higher derivative automorphic forms and string parameter limits}

\vskip 1cm
\centerline{Finn Gubay and Peter West}
\centerline{Department of Mathematics}
\centerline{King's College, London WC2R 2LS, UK}

\vskip .5 cm
\noindent
This paper considers the higher derivative terms in the effective action of type II string theory and in particular the behaviour of the automorphic forms they contain in all the different possible limits of the string parameters. The automorphic forms are thought to obey Poisson equations which contain the Laplacian defined on the coset space to which the scalars fields belong and we  compute this Laplacian in all the possible string theory limits.  We also consider these Poisson equations in  the decompactification limit of a single dimension   and by making two assumptions,  one on the generic form of this equation and the other on the behaviour of the automorphic forms in this limit,  we  find strong constraints on the allowed form of this differential equation. We show that these constraints allow one to recover much of what was previously known about the automorphic forms corresponding to terms in the effective action that have fourteen or fewer space-time derivatives in a simple way. 
\vfill
\eject

\medskip
\noindent
{\bf {1. Introduction}}
\bigskip
The low energy effective actions have played a key role our understanding of strings and branes particularly since there does not exist  an underlying theory of these objects.  of
 For type II string theories these are the maximal supergravity theories in ten dimensions, that is the IIA [1-3] and IIB [4-6] supergravity theories which contain all perturbative and non-perturbative corrections at low energy. The higher space-time derivative corrections  have been studied for many years initially in the context of the IIB theory.   For terms with no more than fourteen space-time derivatives it has been proposed that the coefficients of graviton scattering are  certain SL(2,Z) automorphic forms  that obey Poisson equations [7-15]. These automorphic forms  contain all perturbative and non-perturbative corrections to these terms. Quite a number of these effects have been checked against known string corrections and this provides both strong evidence for these automorphic forms and  also  strong evidence  that the SL(2,R) symmetry of the IIB supergravity theory [4] really is a symmetry of string theory when discretised to SL(2,Z). 
\par
Gravitational higher derivative corrections of type II string theory in lower dimensions, 
and the automorphic forms that might occur,  were discussed quite some time ago [15,16] and was continued in [17,18]. More recently a renewed attempt to understand these correction has been made, specific   automorphic forms have been proposed for the   higher derivative corrections with fourteen and less space-time derivatives. These automorphic forms  have been systematically studied and in particular  their perturbative limits found and shown to agree with string theory results [19-23]. These automorphic forms are constructed from certain representations of $E_{n+1}$ 
where $d=10-n$ is the dimension of the theory. The regularisation of these automorphic forms was also understood [19-23]. 
Although there is some discussion of the the automorphic forms arising as coefficients of terms with more than fourteen space-time derivatives in ten dimensions [14,24,25,26], there has been little discussion of the automorphic forms arising as coefficients of terms with more than fourteen space-time derivatives in less than ten dimensions, with the exception of [12], and the general constraints that were derived in [27, 28]. However, there remains much to be understood about these objects. 
\par
If one knew the automorphic forms that occur in the effective action then one would know all brane and 
string effects, at least for ten and eleven dimensions and their toroidal compactifications. It is likely that one could learn much about the underlying theory of string and branes from these objects. Generally the knowledge that a quantity is some kind of automorphic form places very strong constraints on what this quantity can be. This is familiar to physicists for holomorphic automorphic forms.  The automorphic forms that arise in the higher derivative effective actions are non-holomorphic but instead obey a Poisson equation, that is a Laplace equation that also has a non-zero right-hand side. However, the automorphic forms that are studied in the mathematics literature obey a Laplace equation of the type $\Delta \Phi +\lambda \Phi=0$ where $\Delta $ is the Laplacian on the coset formed by the scalar fields. The automorphic forms that occur for the $R^4$ and $\partial^{4} R^{4}$ terms obey such a Laplace equation, while  the automorphic form for the next correction, $\partial^6R^4$, obeys an equation of the form 
$\Delta \Phi +\lambda \Phi=(\Phi ^{R^4})^2$ where $\Phi^{R^4}$ is the automorphic form for the $R^4$ term [11,20-22]. As a result one can not in general rely on the mathematics literature for help when trying to find the automorphic forms that occur for terms in the effective action that have 
 higher numbers of space-time derivatives. 
\par
An important check on the properties of the automorphic forms that occur for the higher derivative corrections is to study them  as the parameters of string theory are taken to certain limits. The perturbative limit  has been much studied and was used in [19] to provide a powerful check on the acceptability of proposed automorphic forms. In particular most automorphic forms do not lead to perturbative behaviour of the form found in string theory, that is, $g_s^{-2+2n}$ where $g_s$ is the string coupling and $n$ a positive integer. The perturbative limit is achieved by taking the dilaton field $\phi$ to minus infinity as $g_s= e^{\phi}$. The behaviour of automorphic form in this and a number of other limits have been studied in [20,21,22], these include the decompactification of a single dimension limit, the M-theory limit and the $d=10-n$ dimensional perturbative limit. 
\par
String theory in $d=10-n$ dimensions possess $n+2$ parameters. Apart from one dimensional full parameter, which can be taken to be  the Planck length,  the remaining $n+1$ can be thought of as the expectation values of the scalar fields that arise in the non-linear realisation of $E_{n+1}$ from  the Cartan subalgebra of $E_{n+1}$. The string parameters can also be thought of as the Planck length, the string coupling and the parameters of the torus that can be used to find the theory from ten dimensions. This second way of describing  the parameters of string theory has the advantage that taking the various limits corresponds to particular physical processes such a the perturbative limit or certain decompactification limits.  While the former description has the advantage that the expectation values of the scalar fields are closely connected with the group theory used to construct the coset on which the Laplacian mentioned above is defined.  
\par
The relationship between these two  ways of describing the parameters of string theory was given in reference [29]. This paper also contained the precise procedure for taking all the possible the limits, that is,  it specified for each   limit what combination of the fields is taken to a limit and what combination is to be held fixed. In this paper we will use these results to investigate how the Poisson equation behaves in all the possible limits. Since the Poisson equation contains the Laplacian on the scalar coset and the construction of this latter object is found by group considerations. As a result, in this paper  we  use the connections found in reference [29] to find the behaviour of the Laplacian in all possible limits of the string theory parameters, this  is just an exercise in group theory. In section two we recall the relation between the string theory parameters and the parameters used to parameterise the coset group theory element. In section three we give the behaviour of the Laplacian in the possible limits, relying on the results found in appendices A and B,  and we also specify generically how the terms in the effective action behave in these limits. 
\par
In section 4 we consider the Poisson equation satisfied by the automorphic forms in the limit in which one dimension is decompactified. We make two assumptions, one of which concerns the generic form of this equation and the other  the generic behaviour of the automorphic form in this limit. By making these assumptions, and using the results found earlier on the decompactification limit of section 3.4, we find  constraints on the Poisson equation the automorphic form must satisfy. Indeed for the automorphic forms that arise for terms in the effective action with fourteen space-time derivatives or less we are able to completely determine the Poisson equation. Thus from these two assumptions we are able to recover much of what we know about the automorphic forms that occur in the higher derivative corrections to string theory. 
\par
In section 5 we consider the behaviour of the Poisson equation in the perturbative limit and show how it can be used to systematically derive the perturbative behaviour of the above mentioned automorphic forms.  
\par
Given the complexity of the problem of determining the automorphic forms that occur for higher space-time derivatives we hope that starting from our two very natural assumptions will prove a useful way of finding what the automorphic forms can be. 
\medskip
\noindent
{\bf {2. Parameters}}
\bigskip
In this section we review how the parameters arise in string theory as discussed in [29] but we will use a slightly different definition of the parameters.  String theory in $d$ dimensions has $n+2$ parameters where $n=10-d$. These parameters must also occur in the corresponding low energy effective action, that is the maximal supergravity theory in $d$ dimensions. One of these parameters provides the dimensional scale and from the string perspective is the string length $l_s$ while from the supergravity viewpoint it is the Planck length in $d$ dimensions $l_d$ which is related to the  Newtonian coupling $\kappa_d$ by $l_d^{d-2}= 2\kappa_d^{2}$
The remaining $n+1$ dimensionless parameters can  be thought as the expectation values of certain scalars that occur in the supergravity  theories. To give a familiar example;  in the IIA theory in ten dimensions we have two parameters; the Planck length $l_{10 (A)}$ 
and the expectation value of the dilaton as seen from the supergravity viewpoint which corresponds, from the perspective of the string theory, to the string length $l_s$ and the  
string coupling $g_{s (A)}$. 
\par
The scalars in the maximal supergravity theory in $d$ dimensions belong to a non-linear realisation of the group with Lie algebra $E_{n+1}$ and it is the 
expectation values of the scalars associated with the Cartan sub-algebra of $E_{n+1}$ that lead to $n+1$ dimensionless  parameters. We note that unlike the other scalars, the  scalars  associated with the Cartan subalgebra  appear in the supergravity theory as arguments of exponential factors.    Like any semi-simple finite dimensional Lie algebra,  $E_{n+1}$  can be formulated as the multiple commutators of a set of Chevalley generators which include those of the Cartan subalgebra. Indeed, it provides a basis for the Cartan subalgebra, denoted $\{ H_a, a=1,\ldots , n+1\}$,    each generator of which is associated with a  node of the Dynkin diagram of $E_{n+1}$. The part of the group element that occurs in the non-linear realisation and belongs to the Cartan subalgebra can be written in the form $\exp (\sum_a\dot{\varphi}_{a} H_a)$ 
where $\dot{\varphi}_{a}$, $a=1,...,n+1$ are  $n+1$ scalar fields which we will refer to as the Chevalley fields. 
As a result each Chevalley field $\dot{\varphi}_{a}$  can be associated with a node in the $E_{n+1}$ Dynkin diagram. 
We give below the Dynkin diagram of $E_{n+1}$ with the labelling of the nodes which we will use. 
$$
\matrix {
 & & & & & & n+1 & &  &&\cr
 & & & & & & \bullet & & && \cr
 & & & & & & | & & &&\cr
  \bullet & - &
\ldots &-&\bullet&-&\bullet&-&\bullet&-&\bullet \cr
1 & &  & & n-3  &  & n-2 &  & n-1 &&
n}
$$\medskip
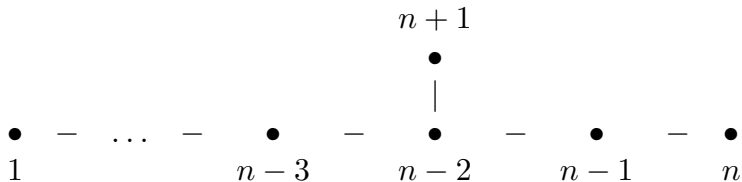
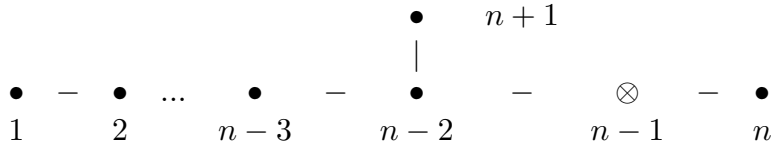
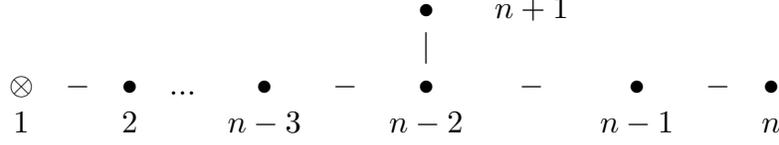
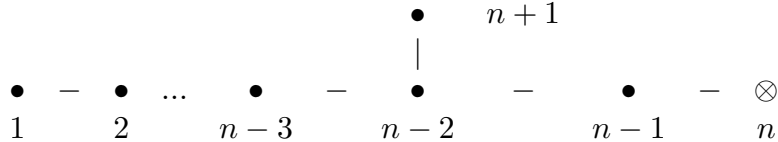
\centerline {The $E_{n+1}$ Dynkin diagram }
\medskip


The  parameters can also be thought to arise from a dimensional reduction process. However, there are three different ways to find the theory in $d$ dimensions by dimensional reduction; we can dimensionally reduce from eleven dimensional M theory on a $n+1$-dimensional torus, the ten dimensional type IIA theory on a $n$-dimensional torus or the IIB theory on a $n$-dimensional torus to find the $d$-dimensional theory. 
As before we can take the one dimensional parameter to be the Planck length in $d$ dimensions $l_d$, but also useful are the Planck length in eleven dimensions, $l_{11}$ and the Planck lengths of the ten dimensional IIA and IIB theories denoted by $l_{10 (A)}$, $l_{10 (B)}$ respectively. Their relations to the corresponding Newton constants are given by the analogue of the above equation for $l_d$, for example 
$l_{11}^{9}=2\kappa_{11}^{2}$. The remaining $n+1$ dimensionless parameters are the volumes  of the $n+1$-dimensional torus, and all its subtori,  used to derive the theory from eleven dimensions. While if we dimensionally reduce from the ten dimensional IIA or IIB theories the $n+1$ parameters are the string coupling $g_{s(A)}$, or $g_{s(B)}$, and the dimensions of the $n$-dimensional torus, and all its subtori. We note that to find the same theory in $d$ dimensions one must use different $n$-dimensional tori when dimensionally reducing from IIA and IIB. Of interest to us is  the relationship between the $n+1$ dimensionless parameters just discussed and the expectation values of the Chevalley fields [29].  

The relation between the expectation values of the Chevalley scalar fields and the above  string couplings are given by [29] 
$$
g_{d}=e^{-\left( {8-n \over 4} \right)\dot{\varphi}_{n}},
\eqno(2.1)
$$
$$
g_{s(A)}=e^{  - {3 \over 2} \dot{\varphi}_{n} + \dot{\varphi}_{n+1}  }
\eqno(2.2)
$$
$$
g_{s(B)}=e^{  - 2 \dot{\varphi}_{n} }
\eqno(2.3)
$$
\par
  Let us  denote the volumes of the respective tori by  
$V_{n+1 (M)}$, $V_{n(A)}$ and  $V_{n(B)}$ which are defined by 
$$
V_{n(A)}= (2 \pi)^{n} {r_{10} r_{9}... r_{d+1} \over (l_{10 (A)})^{n}} =e^{{8-n \over 8} (\dot{\varphi}_{n} + 2 \dot{\varphi}_{n+1})  }
\eqno(2.4)
$$
$$
V_{n(B)}= (2 \pi)^{n} {r_{10} r_{9}... r_{d+1} \over (l_{10 (B)})^{n}} =e^{{ 8-n \over 4} \dot{\varphi}_{n-1}},
\eqno(2.5)
$$
$$
V_{n+1(M)} = (2 \pi)^{n+1} {r_{11} r_{10} r_{9}... r_{d+1} \over (l_{11})^{n+1}}  =e^{{8-n \over 3} \dot{\varphi}_{n+1} },
\eqno(2.6)
$$
In these equations we have also given their expressions in terms of the Chevalley fields. 
\par
For $d<9$ the other $n-1$ parameters that describe the torus can be expressed as the radius of the torus in the $d+1$ direction $r_{d+1}$ 
$$
{r_{d+1} \over l_{d+1} }=e^{{8-n \over 9-n} \dot{\varphi}_{1}},
\eqno(2.7)
$$
the volume of the sub-tori of dimension $n-1$ 
$$
V_{n-1}=(2 \pi)^{n-1} {r_{9}... r_{d+1} \over (l_{9})^{n-1}}=e^{{8-n \over 7}(\dot{\varphi}_{n-1}+\dot{\varphi}_{n+1})},
\eqno(2.8)
$$
and the volumes of the  sub-tori of dimension $j=2,3,...,n-2$ 
$$
V_{j}=(2 \pi)^{j} {r_{d+j} r_{d+j-1} r_{d+j-2}... r_{d+1} \over l_{d+j}^{j}}=e^{{8-n \over 8-n+j} \dot{\varphi}_{j} }, \quad for \ j=2,3,...,n-2.
\eqno(2.9)
$$
\par
Note that  the volumes in equations (2.7-9) are independent of $r_{10}$, that is,  the radius of the torus involved in the dimensional reduction of the type IIA, or type IIB,  theory to nine dimensions, but this quantity appears in the volumes in equations (2.4-6). The 
 remaining radii $r_{9},..., r_{d+1}$ are the  radii of the torus used in the compactification below nine dimensions. 

We refer the reader to [29] for further details. We note that the dimensions of the torus are made dimensionless by dividing by the Planck length, but there is a choice over which Planck length to take. In reference [29] we used the Planck length in $d$ dimensions $l_d$,  but in this paper we have used the Planck length in the decompactified theory  which leads to slightly different expressions in equations (2.4)-(2.9) in terms of the Chevalley fields.  For example the volume of the M-theory torus in reference [29] was defined to be $V_{n+1(M)}=(2\pi)^{n+1} {r_{11}r_{10}...r_{d+1} \over l_{d}^{n+1}}$, whereas in this paper we  take the volume of the M-theory torus to be given by $V_{n+1(M)}=(2\pi)^{n+1} {r_{11}r_{10}...r_{d+1} \over l_{11}^{n+1}}$ in this paper. This also explains why we find the quantity $l_{d+j}$ in equation (2.9) for example.

Clearly, the number of parameters  listed above are more than $n+2$;  the redundancy corresponding to the three different way one can find the $d$-dimensional theory by dimensional reduction of type IIA, type IIB supergravity on an $n$ torus or eleven dimensional supergravity on an $n+1$ torus. The relations between the different parameters are discussed in detail in [29]. We now give the set of  $n+1$ independent dimensionless parameters that arise from the dimensional reduction from eleven dimensions; these are the volume of the $n+1$-dimensional  torus $V_{n+1(M)}$ and its subtori, $V_{1},...,V_{n}$ of equations (2.6)-(2.9). As explained above the Chevalley fields are in one to one correspondence with the nodes of the Dynkin diagram of $E_{n+1}$ in figure 1 and by looking at the expressions for the parameters in terms of these fields we can associate the   parameters with  the nodes of the $E_{n+1}$ Dynkin diagram.  Drawing these on the Dynkin diagram we find that 
$$
\matrix {
 & & & & & & V_{n+1(M)}^{{3 \over 8-n}} & &  &&\cr
 & & & & & & \bullet & & && \cr
 & & & & & & | & & &&\cr
  \bullet & - &
\ldots &-&\bullet&-&\bullet&-&\bullet&-&\bullet \cr
 V_{1}^{{9-n \over 8-n} } & &  & & V_{n-3}^{{5 \over 8-}}  &  & V_{n-2}^{{6 \over 8-n}} &  & V_{n-1}^{{7 \over 8-n}}V_{n+1(M)}^{-{3 \over 8-n}} && V_{n(M)}^{{8 \over 8-n }}V_{n+1(M)}^{-{6 \over 8-n}} }
$$
\bigskip
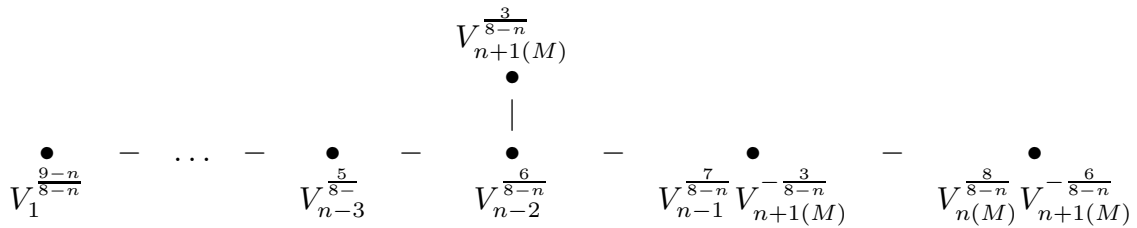
\centerline{Figure 2. The $E_{n+1}$ Dynkin diagram labelled by the $d$
dimensional M-theory parameters}
\bigskip
The meaning of the diagram is that the exponential of the scalar field associated with the node being considered is equal to the quantity shown at that node, for example for node $n+1$ we read off that $ e^{\phi_{n+1}}= 
V_{n+1(M)}^{{3 \over 8-n}}$. having read off all the relations one can express the parameters in terms of the Cartan scalars. 
\par
Similar identifications for the $n+1$ parameters in terms of the dimensional reductions from the type IIA and type IIB theories in reference [29] . 
One may also label the $E_{n+1}$ Dynkin diagram in terms of the parameters resulting from the dimensional reductions of both the type IIA and type IIB theories as shown in figures 2 and 3.  
$$
\matrix {
 & & & & & & V_{n(A)}^{{4 \over 8-n}}g_{d}^{{2 \over 8-n}} & &  &&\cr
 & & & & & & \bullet & & && \cr
 & & & & & & | & & &&\cr
  \bullet & - &
\ldots &-&\bullet&-&\bullet&-&\bullet&-&\bullet \cr
 V_{1}^{{9-n \over 8-n} } & &  & & V_{n-3}^{{5 \over 8-n}}  &  & V_{n-2}^{{6 \over 8-n}} &  & V_{n-1}^{{7 \over 8-n}}V_{n(A)}^{-{4 \over 8-n}}g_{d}^{{4 \over 8-n}} && g_{d}^{-{4 \over 8-n}} }
$$
\bigskip
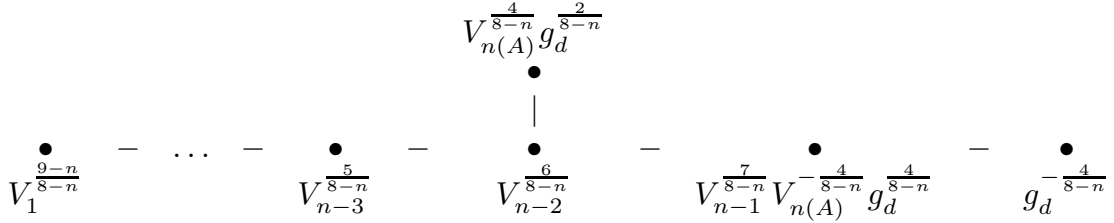
\centerline{Figure 3. The $E_{n+1}$ Dynkin diagram labelled by the $d$
dimensional type IIA parameters}
\bigskip
$$
\matrix {
 & & & & & & V_{n-1}^{{7 \over 8-n}}V_{n(B)}^{{-4 \over 8-n}} & &  &&\cr
 & & & & & & \bullet & & && \cr
 & & & & & & | & & &&\cr
  \bullet & - &
\ldots &-&\bullet&-&\bullet&-&\bullet&-&\bullet \cr
 V_{1}^{{9-n \over 8-n} } & &  & & V_{n-3}^{{5 \over 8-n}}  &  & V_{n-2}^{{6 \over 8-n}} &  & V_{n(B)}^{{4 \over 8-n}} && g_{d}^{-{4 \over 8-n}} }
$$
\bigskip
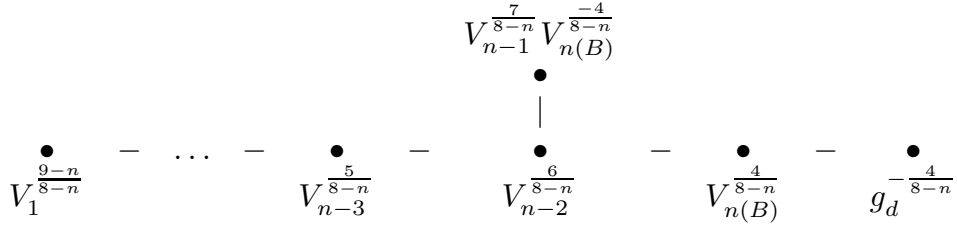
\centerline{Figure 4. The $E_{n+1}$ Dynkin diagram labelled by the $d$
dimensional type IIB parameters}
The relations between the parameters of the $d$ dimensional type IIA, type IIB string theories and M-theory may be derived through the dependence of the parameters on the Chevalley fields, for further details see reference [29]. In the conventions of this paper the relations between the parameters are
$$
V_{n(A)}=V_{n(M)}=V_{n-1}^{{7 \over 4}}g_{d(B)}^{-{1 \over 2}}V_{n(B)}^{-1},
\eqno(2.10)
$$
$$
V_{n(B)}=V_{n+1(M)}^{-{3 \over 4}}V_{n-1}^{{7 \over 4}}=V_{n(A)}^{-1}g_{d(A)}^{-1},
\eqno(2.11)
$$
$$
V_{n+1(M)}=V_{n-1}^{{7 \over 3}}V_{n(B)}^{-{4 \over 3}}=V_{n(A)}^{{4 \over 3}}g_{d(A)}^{-{2 \over 3}},
\eqno(2.12)
$$
$$
V_{n(M)}=V_{n(A)}=V_{n-1}^{{7 \over 3}}g_{d(B)}^{-{1 \over 2}}V_{n(B)},
\eqno(2.13)
$$
$$
g_{d(A)}=g_{d(B)}=V_{n(M)}^{2}V_{n+1(M)}^{3 \over 2},\ \ for\quad  n>0,
\eqno(2.14)
$$
where $g_{d(A)}$ and $g_{d(B)}$ denote the $d=10-n$ dimensional coupling obtained when type IIA or type IIB string theory, respectively, is compactified on an $n$ torus.  Note that the torus subvolumes $V_{j}$, $j=1,...,n-1$, of the type IIA, type IIB and M-theory tori are equivalent, i.e. $V_{k(A)}=V_{k(B)}=V_{k(M)}$, for $k=1,...,n-1$.  


\medskip
\noindent
{\bf {3. Laplacians and automorphic forms in the limits}}
\bigskip
We will be interested in studying the automorphic forms in the limits when certain parameters, or equivalently certain scalar fields, become large or small as appropriate. However, the automorphic forms are thought to satisfy differential equations that contain the Laplacian where differentiation is with respect to the scalar fields of the theory. As such it is useful to study the Laplacian in these limits and in turn  use these results to study the automorphic forms in the limits. Since the automorphic forms are not in general known it is difficult to study their limits, however, one can study the properties they should satisfy in the limits by studying the Poisson equation they satisfy in these limits. This can be used to place restrictions on the automorphic forms. 
\par
The Laplacian lives on the coset space  $ E_{n+1}/H$, where $H$ is the maximal compact subgroup and it can be constructed from a group element 
$ g\in E_{n+1}$ subject to the equivalence relation $g\to gh$ for any group element $h\in H$.  The Laplacian is then given by $\Delta= {1 \over {\sqrt{ \gamma}}} \partial_{i} \left( \sqrt{\gamma} \gamma^{ij} \partial_{j} \right)$ where $\gamma_{ij}$ are the components of the $E_{n+1}/H$ group metric found by tracing over the Cartan forms constructed from $g$.  A full derivation is given in appendix A.
\par
The limits we examine in this paper break the $E_{n+1}$ group into various subgroups. In these limits the Laplacian splits into a Laplacian for the various subgroups plus a part that contains the scalar field being taken to the limit.  In the remainder of this section we present the  behaviour of 
the $E_{n+1}/H$ Laplacians in the various limits as well as the generic behaviour of  higher derivative terms in the $d$ dimensional theory.
\par
The Laplacian in the $d=10-n$ dimensional decompactification of a single dimension limit, large volume limit of the M-theory torus and the perturbative limit was derived in reference [20] by using an iterative method. This method  exploited the fact that the Eisenstein series appearing as the coefficients of the $R^4$ and $\partial^4 R^4$ terms in $d$ dimensions were  known to obey  Laplace  equations in $d=10-n$ dimensions including a knowledge of the  eigenvalues they contain. The form of the Laplacian in these limits was then found by making sure that the known behaviour of the Eisenstein series in these limits did indeed obey the Laplace equations in these limits. In this paper we consider a direct derivation of the Laplacian in these limits 
using its definition in terms of the underlying coset on which it is defined. 
The limits of the $d=10-n$ Laplace operator that we consider include the above limits and in these cases we  agree with the results found in [20].
\par 

\medskip
\noindent
{\bf {3.1 M-theory Limit}}
\bigskip
We begin by studying the large volume limit of the M-theory torus $V_{M(m)}$ in $d=11-m=10-n$ dimensions, that is the decompactification to M theory. In equation (2.6) above we find that  $V_{M(m)}$  is related to the $E_{n+1}$ Chevalley field $\dot{\varphi}_{n+1}$ by $V_{M(m)}= e^{\left({8-n \over 3}\right) \dot{\varphi}_{n+1} }$.  Taking the $V_{M(m)} \rightarrow \infty$ limit is thus the same as taking $\dot{\varphi}_{n+1}$ to the limit and so this breaks the $E_{n+1}$ symmetry leaving a $ GL(1) \times SL(n+1)$ symmetry.  One may think of this as deleting node $n+1$ in the $E_{n+1}$ Dynkin diagram and decomposing the $E_{n+1}$ algebra with respect to the remaining $GL(1) \times SL(n+1)$ subalgebra, for an account of how to carry out this procedure,  see reference [30], however it is important to note that when one takes $V_{M(m)} \rightarrow \infty$ the $E_{n+1}$ symmetry is broken. 
\par
In order to preserve the $SL(n+1)$ symmetry in this limit we find that one must hold fixed the Cartan fields 
$$
\underline{\tilde{\varphi}}=
 \sum_{a=1}^{n-1}{\dot \varphi}_a \underline{\alpha}_{a} - \dot{\varphi}_{n} \underline{\lambda}_{n-1}+ \dot{\varphi}_{n+1} \underline{\alpha}_{n} 
\eqno(3.1.1)
$$
where $\underline{\alpha}_{i}$ and $\underline{\lambda}_{i}$, $i=1,...,n$ are the simple roots and fundamental weights of $SL(n+1)$.  We refer the reader to section 4.1.4 of reference [29] for a detailed discussion of this point. 
\par
In the large volume limit of the M-theory torus $V_{M(m)}= e^{-\left({8-n \over 3}\right) \dot{\varphi}_{n+1} } \rightarrow \infty$ the Laplacian $\Delta$ becomes
$$
\Delta= { 1 \over 2x^{2} } { \partial \over \partial \dot{\varphi}_{n+1} } { \partial \over \partial \dot{\varphi}_{n+1} } - { (3n^{2} -n - 4) \over 2(8-n)   } { \partial \over \partial \dot{\varphi}_{n+1} } + \Delta_{SL(n+1)}
$$
$$
= { n+1 \over 2(8-n) } { \partial \over \partial \dot{\varphi}_{n+1} } { \partial \over \partial \dot{\varphi}_{n+1} } - { (3n^{2} -n - 4) \over 2(8-n)   } { \partial \over \partial \dot{\varphi}_{n+1} } + \Delta_{SL(n+1)}
\eqno(3.1.2)
$$
where we have used $x^{2}= {8-n \over n+1}$. We refer the reader to appendix B.1 for a detailed derivation
\par
By dimensional analysis one sees that an arbitrary $d$ dimensional higher derivative term in Einstein frame that occurs in the effective action takes the form 
$$
l_{d}^{k-d}  \int d^{d}x \sqrt{-g}\Phi_{E_{n+1}} {\cal{O}}
\eqno(3.1.3)
$$
where ${\cal{O}}$ is a $k$ derivative polynomial in the $d$ dimensional curvature
$R$, Cartan forms $P$ or field strengths $F$. 
\par
We now examine how the automorphic form in equation (3.1.3)  behaves in the large volume limit of the M-theory torus. To do this   we will convert the $d$ dimensional Planck length $l_{d}$ to the eleven dimensional Planck length $l_{11}$ and the volume of the M-theory torus $V_{M(m)}$ using the relation 
$$
l_{d}=l_{11} V_{M(m)}^{-{1 \over 8-n}}
\eqno(3.1.4)
$$
and the condition
$$
\lim_{V_{m(M)} \rightarrow \infty} l_{11}^{n+1} \int d^{d} x \sqrt{-g} V_{m(M)} = \int d^{11}x \sqrt{-\hat{g}}, 
\eqno(3.1.5)
$$
Applying this limit to the general term of equation (3.1.3) we conclude that any term which is linear in $V_{m(M)}$  is preserved in the limit while any term with a power of $V_{m(M)}^p$ for $p<1$ vanishes in the limit.  
\par
In the $V_{m(M)} \rightarrow \infty$ limit the $E_{n+1}$ coefficient function $\Phi_{E_{n+1}}$ generically splits as
$$
\Phi_{E_{n+1}}= \sum_{i} V_{M(m)}^{ a_{i}} \Phi_{SL(n+1)}^{i}
\eqno(3.1.6)
$$
where $i$ labels the different $SL(n+1)$ coefficient functions, that is,  $SL(n+1)$ automorphic forms, arising in the limit and $a_{i}$ is a real number.   Demanding that the large volume limit of this generic higher derivative term converges to an acceptable higher derivative term in the M-theory effective action implies  that the large volume limit $V_{m(M)} \rightarrow \infty$ exists and that the resulting terms in  eleven dimensions have  constant coefficients rather than non-trivial  $SL(n+1)$ automorphic forms. Put another way the eleven dimensional terms in the M theory effective action can not depend on the moduli of the torus. We note that an  $SL(n+1)$ which is built from the trivial representation is a constant.   Using equations (3.1.4) and (3.1.5) and the  decomposition of the automorphic form of equation (3.1.6) we find the generic term of equation (3.1.3) can be written in the limit in the form 
$$
l_{d}^{k-d}  \int d^{d}x \sqrt{-g}\Phi_{E_{n+1}} {\cal{O}} = l_{11}^{k-11}  \int d^{11}x \sqrt{-\hat{g}} \lim_{V_{m(M)} \rightarrow \infty} V_{M(m)}^{{2-k \over 8-n}} (\sum_{i} V_{M(m)}^{ a_{i}} \Phi_{SL(n+1)}^{i} ) {\cal{O}}
$$
$$
=l_{11}^{k-11}  \int d^{11}x \sqrt{-\hat{g}} b \hat{{\cal{O}}}
\eqno(3.1.7)
$$
where $\hat{\cal{O}}$ denotes  the different $d=11$ M-theory polynomials in the eleven dimensional curvature $\hat{R}$, and field strengths $\hat{F}$ that arise in the decompactification of the $d$ dimensional polynomial in the curvature $R$, Cartan forms $P$ and field strengths $F$. We have in the last line of equation (3.1.7) encoded the requirement, mentioned above,   that the only $SL(n+1)$ coefficient functions that can be preserved in the limit are constants,  denoted by $b$.  The terms that are clearly preserved in this limit are those in $\sum_{i} V_{M(m)}^{ a_{i}} \Phi_{SL(n+1)}^{i} $ with $V_{M(m)}^{-\left({2-k \over 8-n} \right)}$ as  in this case the factor of $V_{M(m)}$ combines with that contributed from converting the $d$ dimensional Planck length to the eleven dimensional Planck and $V_{M(m)}$ via equation (3.1.4) to converge to an eleven dimensional higher derivative term.  Terms with a lesser power of $V_{M(m)}$ vanish in the $V_{M(m)} \rightarrow \infty$, while those with a greater power are non-analytic and must be treated carefully. We refer to the references [12,20,21,22,24,26,31] for a discussion of this point.

Having found the terms that result in the decompactification to eleven dimensions we can demand that they  match the known coefficient functions of the higher derivative terms in the M-theory effective action in eleven dimensions. As we will demonstrate in section 4 we can  apply the Laplacian when written in the limit to the automorphic form when also written in the limit and for certain limits this can place strong constraints on the form of the differential equation satisfied by the automorphic form and as a result the automorphic form itself. 

\medskip
\noindent
{\bf {3.2 Perturbative Limit}}
\bigskip
The string coupling $g_{d}$ in $d=10-n$ dimensions is related to the $E_{n+1}$ Chevalley field $\dot{\varphi}_{n}$ by 
$g_{d}= e^{-\left({8-n \over 4}\right) \dot{\varphi}_{n} }$.  Taking the $g_{d} \rightarrow 0$ limit is the same as taking $\dot{\varphi}_{n}\to \infty$ and so it breaks the $E_{n+1}$ symmetry leaving a $GL(1) \times SO(n,n)$ symmetry.  One may think of this as deleting node $n$ in the $E_{n+1}$ Dynkin diagram and decomposing the $E_{n+1}$ algebra with respect to the remaining $GL(1) \times SO(n,n)$ subalgebra, for an account of how to carry out this procedure in general see reference [30]. It is important to note that when one takes $g_{d} \rightarrow 0$ the $E_{n+1}$ symmetry is broken to $GL(1) \times SO(n,n)$ .  
\par
In order to preserve the $SO(n,n)$ symmetry in the perturbative limit $g_{d} \rightarrow 0$ we find that one must hold fixed the Cartan fields 
$$\underline{\tilde{\varphi}}=
 \sum_{a=1}^{n-1}{\dot \varphi}_a \underline{\alpha}_{a} - \dot{\varphi}_{n} \underline{\lambda}_{n-1}+ \dot{\varphi}_{n+1} \underline{\alpha}_{n}, 
\eqno(3.2.1)
$$
where $\underline{\alpha}_{i}$ and $\underline{\lambda}_{i}$, $i=1,...,n-1$ are the simple roots and fundamental weights of $SO(n,n)$ respectively. We refer the reader to section 4.1.2 of reference [29] for a detailed discussion of this point.
\par
In the perturbative limit $g_{d}= e^{-\left({8-n \over 4}\right) \dot{\varphi}_{n} } \rightarrow 0$ the Laplacian $\Delta$ becomes
$$
\Delta= { 1 \over 2x^{2} } { \partial \over \partial \dot{\varphi}_{n} } { \partial \over \partial \dot{\varphi}_{n} } - { (n^{2} -n + 4) \over (8-n)   } { \partial \over \partial \dot{\varphi}_{n} } + \Delta_{SO(n,n)}
$$
$$
= { 4 \over 2(8-n) } { \partial \over \partial \dot{\varphi}_{n} } { \partial \over \partial \dot{\varphi}_{n} } - { (n^{2} -n + 4) \over (8-n)   } { \partial \over \partial \dot{\varphi}_{n} } + \Delta_{SO(n,n)}
\eqno(3.2.2)
$$
where we have used $x^{2}= {8-n \over 4}$
\par
We require that the perturbative terms are consistent with a
perturbative expansion in $g_{d}$.  In string frame this implies that
each term has a $g_d$ dependence that is of the form
$g_{d}^{2g-2}$ where
$g$ is the genus.  String frame in $d$ dimensions is related to Einstein
frame by \
$g_{E \mu\nu}=g_{d}^{-{4 \over d-2}}g_{S \mu \nu}$. Upon rescaling to string
frame, an arbitrary higher derivative term in the $d=10-n$ dimensional type II string theory effective action, as given in equation (3.1.3),  is given by
$$
\int d^{d}x \sqrt{-g_{S}}g_{d}^{{4\Delta-2d \over d-2}}\Phi_{E_{n+1}}
{\cal{O}}_{S},
\eqno(3.2.3)
$$
where ${\cal{O}}_S$  is some polynomial in the $d$ dimensional curvature
$R$, Cartan forms $P$,  or field strengths $F$, the subscript $S$ denotes
string frame quantities and $\Delta$ is the number of space time metrics
minus the number of inverse space time
metrics in ${\cal{O}}_{S}$.
Demanding that the perturbative limit of  this generic higher derivative
term exists from a string theory perspective means that in the limit
$g_{d} \rightarrow 0$ any term in the effective action agrees with a perturbative
expansion in $g_{d}$,  for this one requires that each term is
multiplied
by a factor of the form $g_{d}^{-2+2n}$, where $n$ is either zero or a
positive integer. Given a putative  automorphic form we can compute its behaviour in the perturbative limit and having substituted this into equation (3.2.3) we can test if it has an acceptable string perturbation theory or not. This was indeed what was done in reference [19,20-22] and it was found to be a very restrictive requirement.

\medskip
\noindent
{\bf {3.3 Type IIB Limit}}
\bigskip
The volume of the type IIB torus $V_{n(B)}$ in $d=10-n$ dimensions is related to the $E_{n+1}$ Chevalley field $\dot{\varphi}_{n-1}$ by $V_{n(B)}= e^{{8-n \over 4} \dot{\varphi}_{n-1} }$.  Taking the $V_{n(B)} \rightarrow \infty$ limit corresponds to taking $\dot{\varphi}_{n-1}\to \infty$  and so it breaks the $E_{n+1}$ symmetry leaving a $GL(1) \times SL(2) \times SL(n)$ symmetry.  One may think of this as deleting node $n-1$ in the $E_{n+1}$ Dynkin diagram and decomposing the $E_{n+1}$ algebra with respect to the remaining $GL(1) \times SL(2) \times SL(n)$ subalgebra. 
\par
In order to preserve the $SL(2) \times SL(n)$ symmetry in this limit we find that one must hold fixed the Cartan fields 
$$
\underline{\tilde{\varphi}}=
\sum_{a=1}^{n-2}{\dot \varphi}_a \underline{\alpha}_{a} + \dot{\varphi}_{n+1} \underline{\alpha}_{n-1} - \dot{\varphi}_{n-1} \underline{\lambda}_{n-2} 
\eqno(3.3.1)
$$
where $\underline{\alpha}_{i}$ and $\underline{\lambda}_{i}$, $i=1,...,n-1$ are the simple roots and fundamental weights of $SL(n)$ respectively and in addition fix
$$
\tilde{\varphi}=\mu\dot{\varphi}_{n-1}- \beta \dot{\varphi}_{n}
\eqno(3.3.2)
$$
where $\mu$ and $\beta$ are fundamental weight and simple root of $SL(2)$. We refer the reader to section 4.1.2 of reference [29] for a detailed discussion of this point. 
\par
In the large volume limit of the type IIB torus $V_{n(B)}= e^{{8-n \over 4} \dot{\varphi}_{n-1} } \rightarrow \infty$ the Laplacian $\Delta$ becomes
$$
\Delta= { 1 \over 2x^{2} } { \partial \over \partial \dot{\varphi}_{n-1} } { \partial \over \partial \dot{\varphi}_{n-1} } - { (2n^{2} - n ) \over (8-n)   } { \partial \over \partial \dot{\varphi}_{n-1} } + \Delta_{SL(n)} + \Delta_{SL(2)}
$$
$$
= { n \over (8-n) } { \partial \over \partial \dot{\varphi}_{n-1} } { \partial \over \partial \dot{\varphi}_{n-1} } - { (2n^{2} - n ) \over (8-n)   }  { \partial \over \partial \dot{\varphi}_{n-1} } + \Delta_{SL(n)}+ \Delta_{SL(2)}
\eqno(3.3.3)
$$
where we have used $x^{2}= {8-n \over 2n}$. A partial derivation is given in appendix B.2. 
\par
The generic higher derivative term in the effective action was given in equation (3.1.3) and we now examine how the automorphic form behaves in the large volume limit of the type IIB torus.   To do this  we will convert the $d$ dimensional Planck length $l_{d}$ to the type IIB ten dimensional Planck length $l_{10 (B)}$ and the volume of the type IIB torus $V_{n(B)}$ using the relation 
$$
l_{d}=l_{10(B)} V_{n(B)}^{-{1 \over 8-n}}
\eqno(3.3.4)
$$
and use the condition
$$
\lim_{V_{n(B)} \rightarrow \infty} l_{10(B)}^{n} \int d^{d} x \sqrt{-g} V_{n(B)} = \int d^{10}x \sqrt{-\hat{g}}, 
\eqno(3.3.5)
$$
which implies that any term with $V_{n(B)}$ dependence $V_{n(B)}$ is preserved in the limit while any term with a lesser power of $V_{n(B)}$ vanishes in the limit.  In the $V_{n(B)} \rightarrow \infty$ limit the $E_{n+1}$ symmetry is broken, leaving a $GL(1) \times SL(2)  \times SL(n)$ symmetry.  The $E_{n+1}$ coefficient function $\Phi_{E_{n+1}}$ generically splits into an expansion in the volume of the type IIB torus $V_{n(B)}$ as
$$
\Phi_{E_{n+1}}= \sum_{i} V_{n(B)}^{ a_{i}} \Phi_{SL(2)}^{i} \Phi_{SL(n)}^{i} 
\eqno(3.3.6)
$$
where $i$ labels the different $SL(2)$ and $SL(n)$ coefficient functions arising in the limit and $a_{i}$ is a real number.   Demanding that the large volume limit of this generic higher derivative term converges to an acceptable higher derivative term in the type IIB effective action means that the large volume limit $V_{n(B)} \rightarrow \infty$ exists and that the resulting terms are ten dimensional type IIB higher derivative terms have  coefficient functions that are only $SL(2)$ automorphic forms and that the $SL(n)$  automorphic forms become constants in the limit since  the ten dimensional type IIB effective action can not depend on the moduli of the $n$ torus.  This condition may be expressed as
$$
l_{d}^{k-d}  \int d^{d}x \sqrt{-g}\Phi_{E_{n+1}} {\cal{O}} = l_{10 (B)}^{k-10}  \int d^{10}x \sqrt{-\hat{g}} \lim_{V_{n(B)} \rightarrow \infty} V_{n(B)}^{{2-k \over 8-n}} (\sum_{i} V_{n(B)}^{ a_{i}} \Phi_{SL(2)}^{i} \Phi_{SL(n)}^{i} ) {\cal{O}}
$$
$$
=l_{10 (B)}^{k-10}  \int d^{10}x \sqrt{-\hat{g}} (\sum_{i} b_{i} \Phi_{SL(2)}^{i} ) \hat{{\cal{O}}}
\eqno(3.3.7)
$$
where $b_{i}$ is a real number and $\hat{\cal{O}}$ labels the different $d=10$ type IIB polynomials in the ten dimensional curvature $\hat{R}$, and field strengths $\hat{F}$ that arise in the decompactification of the $d$ dimensional polynomial in the curvature $R$, Cartan forms $P$ and field strengths $F$.  The last line of equation (3.3.7) encodes the requirement that we find only $SL(2)$ automorphic forms  in  the ten dimensional type IIB theory.  The terms that are clearly preserved in this limit are those in $\sum_{i} V_{n(B)}^{ a_{i}}  \Phi_{SL(2)}^{i} \Phi_{SL(n)}^{i} $ with $V_{n(B)}^{-\left({2-k \over 8-n} \right)}$, in this case the factor of $V_{n(B)}$ combines with that contributed from converting the $d$ dimensional Planck length to the type IIB ten dimensional Planck length and $V_{n(B)}$ via equation (3.3.4) to converge to a ten dimensional type IIB higher derivative term.  Terms with a lesser power of $V_{n(B)}$ vanish in the $V_{n(B)} \rightarrow \infty$, while those with a greater power are non-analytic and must be treated carefully.

\medskip
\noindent
{\bf {3.4 Decompactification of a single dimension limit}}
\bigskip
The ratio of the radius in the $d+1$ direction $r_{d+1}$ to the $d+1$ dimensional Planck length $l_{d+1}$ in $d=10-n$ dimensions is related to the $E_{n+1}$ Chevalley field $\dot{\varphi}_{1}$ by ${r_{d+1} \over l_{d+1} }= e^{{8-n \over 9-n} \dot{\varphi}_{1} }$.  Taking the ${r_{d+1} \over l_{d+1} } \rightarrow \infty$ limit corresponds to taking $\dot{\varphi}_{1}\to \infty$ and so corresponds to breaks the $E_{n+1}$ symmetry leaving a $GL(1) \times E_{n}$ symmetry.  One may think of this as deleting node $1$ in the $E_{n+1}$ Dynkin diagram and decomposing the $E_{n+1}$ algebra with respect to the remaining $GL(1) \times E_{n}$ subalgebra. 
\par
In order to preserve the $E_{n}$ symmetry in this limit we find that one must hold fixed the Cartan fields 
$$\underline{\tilde{\varphi}}=
\sum_{a=2}^{n+1}{\dot \varphi}_a \underline{\alpha}_{a-1} - \dot{\varphi}_{1} \underline{\lambda}_{1}
\eqno(3.4.1)
$$
where $\underline{\alpha}_{i}$ and $\underline{\lambda}_{i}$, $i=1,...,n-1$ are the simple roots and fundamental weights of $E_{n}$ respectively.   We refer the reader to section 4.1.3 of reference [29] for a detailed discussion of this point.
\par
In the decompactification of a single dimension limit ${r_{d+1} \over l_{d+1} }= e^{{8-n \over 9-n} \dot{\varphi}_{1} } \rightarrow \infty$ the Laplacian $\Delta$ becomes
$$
\Delta= { 1 \over 2x^{2} } { \partial \over \partial \dot{\varphi}_{1} } { \partial \over \partial \dot{\varphi}_{1} } - { (-n^{2} + 17n - 12) \over (2x^{2}(9-n))   } { \partial \over \partial \dot{\varphi}_{1} } + \Delta_{E_{n}}
$$
$$
= { (9-n) \over 2(8-n) } { \partial \over \partial \dot{\varphi}_{1} } { \partial \over \partial \dot{\varphi}_{1} } - { (-n^{2} + 17n - 12) \over 2(8-n)   } { \partial \over \partial \dot{\varphi}_{1} } + \Delta_{E_{n}}
\eqno(3.4.2)
$$
where we have used $x^{2}= {8-n \over 9-n}$. We refer to appendix B.4 for a partial derivation of this result. This result was found in [20] by testing out a putative Laplacian on the Eisenstein series automorphic forms that were know to obey the Laplace equation. Our approach is in this paper is quite different in that we derived the above result in group theory and we are going use it to place restrictions on the automorphic forms that occur in string theory. 
\par
We now examine how the automorphic form behaves in the decompactification of a single dimension limit.  We require that the terms remaining in the decompactification of a single dimension limit match the known coefficient functions of the higher derivative terms in the type II effective action in $d+1$ dimensions. 

The generic higher derivative term in the effective action was given in equation (3.1.3) and we now examine the behaviour of this term in the decompactification of a single dimension limit. To do this  we will convert the $d$ dimensional Planck length $l_{d}$ to the $d+1$ dimensional Planck length $l_{d+1}$ and the ratio of the radius in the $d+1$ direction $r_{d+1}$ to the $d+1$ dimensional Planck length $l_{d+1}$ using the relation 
$$
l_{d}=l_{d+1} \left( {r_{d+1} \over l_{d+1}} \right)^{-{1 \over 8-n}}
\eqno(3.4.3)
$$
and use the condition
$$
\lim_{{r_{d+1} \over l_{d+1}} \rightarrow \infty}  l_{d+1} \int d^{d}x \sqrt{-g} {r_{d+1} \over l_{d+1}}= \int d^{d+1}x \sqrt{-\hat{g}}, 
\eqno(3.4.4) 
$$
which implies that any term which is linear in ${r_{d+1} \over l_{d+1}}$ is preserved in the limit while any term with $\left({r_{d+1} \over l_{d+1}}\right)^{p}$ for $p<1$ vanishes in the limit.  In the ${r_{d+1} \over l_{d+1}} \rightarrow \infty$ limit the $E_{n+1}$ symmetry is broken to a $GL(1) \times E_{n}$ symmetry and the $E_{n+1}$ coefficient function $\Phi_{E_{n+1}}$ generically splits into an expansion in the ratio of the radius in the $d+1$ direction to the $d+1$ dimensional Planck length ${r_{d+1} \over l_{d+1}}$ as
$$
\Phi_{E_{n+1}}= \sum_{i} {\left({r_{d+1} \over l_{d+1}}\right)}^{ a_{i}}  \Phi_{E_{n}}^{i} 
\eqno(3.4.5)
$$
where $i$ labels the different $E_{n}$ coefficient functions arising in the limit and $a_{i}$ is a real number.   Demanding that the large volume limit of this generic higher derivative term converges to an acceptable higher derivative term in the $d+1$ dimensional effective action of type II string theory means that the decompactification of a single dimension limit ${r_{d+1} \over l_{d+1}} \rightarrow \infty$ exists and that the resulting terms are $d+1$ dimensional higher derivative terms in the type II string theory effective action with coefficient functions that are $E_{n}$ automorphic forms.  This condition may be expressed as
$$
l_{d}^{k-d}  \int d^{d}x \sqrt{-g}\Phi_{E_{n+1}} {\cal{O}} = l_{d+1}^{k-(d+1)}  \int d^{d+1}x \sqrt{-\hat{g}} \lim_{{r_{d+1} \over l_{d+1}} \rightarrow \infty} {r_{d+1} \over l_{d+1}}^{{2-k \over 8-n}} (\sum_{i} {r_{d+1} \over l_{d+1}}^{ a_{i}} \Phi_{E_{n}}^{i} ) {\cal{O}}
$$
$$
={ l_{d+1}}^{k-(d+1)}  \int d^{d+1}x \sqrt{-\hat{g}} (\sum_{i}  \Phi_{E_{n}}^{i} ) \hat{{\cal{O}}}
\eqno(3.4.6)
$$
where $\hat{\cal{O}}$ labels the different $d+1$ dimensional type II string theory polynomials in the $d+1$ dimensional curvature $\hat{R}$, and field strengths $\hat{F}$ that arise in the decompactification of the $d$ dimensional polynomial in the curvature $R$, Cartan forms $P$ and field strengths $F$.  The last line of equation (3.4.6) expresses the requirement that the only allowed coefficient functions in the $d+1$ dimensional type II string theory effective action are $E_{n}$ automorphic forms.  The terms that are clearly preserved in this limit are those in $\sum_{i} {r_{d+1} \over l_{d+1}}^{ a_{i}} \Phi_{E_{n}}^{i}  $ with ${r_{d+1} \over l_{d+1}}^{-\left({2-k \over 8-n} \right)}$, in this case the factor of $V_{n(B)}$ combines with that contributed from converting the $d$ dimensional Planck length to the $d+1$ dimensional Planck length and ${r_{d+1} \over l_{d+1}}$ via equation (3.4.3) to converge to a $d+1$ dimensional higher derivative term.  Terms with a lesser power of ${r_{d+1} \over l_{d+1}}$ vanish in the ${r_{d+1} \over l_{d+1}} \rightarrow \infty$, while those with a greater power are non-analytic and must be treated carefully.

\medskip
\noindent
{\bf {3.5 Decompactification of a $j$ dimensional subtorus limit}}
\bigskip
The $j$ dimensional subtorus $V_{j}$ of an $n$ torus in $d=10-n$ dimensions is related to the $E_{n+1}$ Chevalley field $\dot{\varphi}_{j}$ by $V_{j}= e^{{8-n \over 8-n+j} \dot{\varphi}_{j} }$.  Taking the $V_{j} \rightarrow \infty$ limit breaks the $E_{n+1}$ symmetry leaving a $GL(1) \times SL(j) \times E_{n+1-j}$ symmetry.  One may think of this as deleting node $j$ in the $E_{n+1}$ Dynkin diagram and decomposing the $E_{n+1}$ algebra with respect to the remaining $GL(1) \times SL(j) \times E_{n+1-j}$ subalgebra. 
\par
In order to preserve the $SL(j) \times E_{n+1-j}$ symmetry in this limit we find that one must hold fixed the Cartan fields
$$
\tilde{\underline{\varphi}} =\sum_{i=1}^{j-1}\dot{\varphi}_{i} \underline{\alpha}_{i} - \dot{\varphi}_{j} \underline{\lambda}_{j-1}
\eqno(3.5.1)
$$
and $n+1-j$ quantities 
$$
{\hat{\varphi}} =\sum_{a=j+1}^{n+1}{\dot \varphi}_{a} \hat{\alpha}_{a-j} - \dot{\varphi}_{j} \hat{\lambda}_{1}
\eqno(3.5.2)
$$ 
to preserve the $SL(j) \times E_{n+1-j}$ symmetry where $\underline{\alpha}_{i}$ and $\underline{\lambda}_{i}$, $i=1,...,j-1$ are the simple roots and fundamental weights of $SL(j)$ and $\hat{\alpha}_{i}$ and $\hat{\lambda}_{i}$, $i=1,...,n+1-j$ are the simple roots and fundamental weights of $E_{n+1-j}$.  We refer the reader to section 4.1.6 of reference [29] for a detailed discussion of this point. 
\par
In the large volume limit of the $j$ dimensional subtorus $V_{j}= e^{{8-n \over 8-n+j} \dot{\varphi}_{j} } \rightarrow \infty$ the Laplacian $\Delta$ becomes
$$
\Delta= { 1 \over 2x^{2}  } { \partial \over \partial \dot{\varphi}_{j} } { \partial \over \partial \dot{\varphi}_{j} } - { (-n^{2} + 16n - 8j +nj - 4) \over 2x^{2}(8-n+j)   } { \partial \over \partial \dot{\varphi}_{j} } + \Delta_{E_{n+1-j}}
$$
$$
= { j(n+1-j)(8-n+j) \over 2((n+1)(8-n+j) -9j) } { \partial \over \partial \dot{\varphi}_{j} } { \partial \over \partial \dot{\varphi}_{j} } 
$$
$$
- { j(n+1-j) \over 2((n+1)(8-n+j) -9j) } (-n^{2}+16n-8j+nj-4) { \partial \over \partial \dot{\varphi}_{j} } + \Delta_{E_{n+1-j}}
\eqno(3.5.3)
$$
where we have used $x^{2}= {(n+1)(8-n+j) - 9j \over j(n+1-j)(8-n+j)}$.
\par
The generic term in the higher derivative action in $d$ dimensions was given in equation (3.1.3) and we now examine how the automorphic form generically behaves in the large volume limit of a $j$ dimensional subtorus.  We require that the terms remaining in the large volume limit of the $j$ dimensional subtorus match the known coefficient functions of the higher derivative terms in the type II string effective action in $d+j$ dimensions. 
 To examine the behaviour of such terms in the large volume limit of $j$ dimensional  subtorus we will convert the $d$ dimensional Planck length $l_{d}$ to the $d+j$ dimensional Planck length $l_{d+j}$ and the volume of the $j$ dimensional subtorus $V_{j}$ using the relation 
$$
l_{d}=l_{d+j} V_{j}^{-{1 \over 8-n}}
\eqno(3.5.4)
$$
and use the condition
$$
\lim_{V_{j} \rightarrow \infty} l_{d+j}^{j} \int d^{d} x \sqrt{-g} V_{j} = \int d^{d+j}x \sqrt{-\hat{g}}, 
\eqno(3.5.5)
$$
which implies that any term linear in $V_{j}$ is preserved in the limit while any term with a lesser power of $V_{j}$ vanishes in the limit.  In the $V_{j} \rightarrow \infty$ limit the $E_{n+1}$ symmetry breaks into a $GL(1) \times SL(j)  \times E_{n+1-j}$ symmetry and the $E_{n+1}$ coefficient function $\Phi_{E_{n+1}}$ generically splits into an expansion in the volume of the $j$ dimensional subtorus $V_{j}$ as
$$
\Phi_{E_{n+1}}= \sum_{i} V_{j}^{ a_{i}} \Phi_{SL(j)}^{i} \Phi_{E_{n+1-j}}^{i} 
\eqno(3.5.6)
$$
where $i$ labels the different $SL(j)$ and $E_{n+1-j}$ coefficient functions arising in the limit and $a_{i}$ is a real number.   Demanding that the large volume limit of this generic higher derivative term converges to an acceptable higher derivative term in the $d+j$ dimensional type II string theory effective action means that the large volume limit $V_{j} \rightarrow \infty$ exists and that the resulting terms are $d+j$ dimensional type II string theory higher derivative terms with coefficient functions that are $E_{n+1-j}$ automorphic forms. In other words the 
$SL(j)$ automorphic forms only lead to constants in the limit since the $d+j$ dimensional type II string theory effective action can not depend on the moduli of the $j$ dimensional subtorus.  This condition may be expressed as
$$
l_{d}^{k-d}  \int d^{d}x \sqrt{-g}\Phi_{E_{n+1}} {\cal{O}} = l_{d+j}^{k-(d+j)}  \int d^{d+j}x \sqrt{-\hat{g}} \lim_{V_{j} \rightarrow \infty} V_{j}^{{2-k \over 8-n}} (\sum_{i} V_{j}^{ a_{i}} \Phi_{SL(j)}^{i} \Phi_{E_{n+1-j}}^{i} ) {\cal{O}}
$$
$$
=l_{d+j}^{k-(d+j)}  \int d^{d+j}x \sqrt{-\hat{g}} ( \sum_{i} b_{i} \Phi_{E_{n+1-j}}^{i}) \hat{{\cal{O}}}
\eqno(3.5.7)
$$
where $b_{i}$ is a real number and $\hat{\cal{O}}$ labels the different $d=10$ type IIB polynomials in the ten dimensional curvature $\hat{R}$, and field strengths $\hat{F}$ that arise in the decompactification of the $d$ dimensional polynomial in the curvature $R$, Cartan forms $P$ and field strengths $F$.  The last line of equation (3.5.7) expresses the above requirement that the only allowed coefficient functions in the $d+j$ dimensional type II string theory effective action are $E_{n+1-j}$ automorphic forms.   The terms that are clearly preserved in this limit are those in $\sum_{i} V_{j}^{ a_{i}} \Phi_{SL(j)}^{i} \Phi_{E_{n+1-j}}^{i} $ with $V_{j}^{-\left({2-k \over 8-n} \right)}$, in this case the factor of $V_{j}$ combines with that contributed from converting the $d$ dimensional Planck length to the $d+j$ dimensional Planck length and $V_{j}$ via equation (3.5.4) to converge to a $d+j$ dimensional higher derivative term.  Terms with a lesser power of $V_{j}$ vanish in the $V_{j} \rightarrow \infty$ limit, while those with a greater power are non-analytic and must be treated carefully.

\medskip
\noindent
{\bf {3.6 Type IIA Limit}}
\bigskip
The volume of the type IIA torus $V_{n(A)}$ in $d=10-n$ dimensions is related to the $E_{n+1}$ Chevalley fields $\dot{\varphi}_{n}$ and $\dot{\varphi}_{n+1}$ by $V_{n(A)}= e^{{8-n \over 8} (\dot{\varphi}_{n}+2\dot{\varphi}_{n+1}) }$ in addition the ten dimensional type IIA string coupling $g_{s(A)}$ is related to the Chevalley fields by $g_{s(A)}=e^{-{3 \over 2}\dot{\varphi}_{n}+2\dot{\varphi}_{n+1} }$.  Taking the $V_{n(A)} \rightarrow \infty$ limit corresponds to taking $-{3 \over 2}\dot{\varphi}_{n}+2\dot{\varphi}_{n+1} \to\infty$ and it breaks the $E_{n+1}$ symmetry leaving a $GL(1) \times GL(1) \times SL(n)$ symmetry.  One may think of this as deleting nodes $n$ and $n+1$ in the $E_{n+1}$ Dynkin diagram and decomposing the $E_{n+1}$ algebra with respect to the remaining $GL(1) \times GL(1) \times SL(n)$ subalgebra. 
\par
In order to preserve the $SL(n)$ symmetry in this limit we find that one must hold fixed the Cartan fields 
$$
\tilde{\underline{\varphi}}=\sum_{a=1}^{n-1}{\dot \varphi}_a \underline{\alpha}_{a} - \dot{\varphi}_{n} \underline{\lambda}_{n-1} - \dot{\varphi}_{n+1} \underline{\lambda}_{n-2}
\eqno(3.6.1)
$$
where $\underline{\alpha}_{i}$ and $\underline{\lambda}_{i}$, $i=1,...,n-1$ are the simple roots and fundamental weights of $SL(n)$ respectively and in addition fix 
$$
\varphi_{g}=-{3 \over 2} \dot{\varphi}_{n} + \dot{\varphi}_{n+1}
\eqno(3.6.2)
$$
to preserve the type IIA string coupling.   We refer the reader to section 4.1.5 of reference [29] for a detailed discussion of this point. 
\par
In the large volume limit of the type IIA torus $V_{n(A)}= e^{{8-n \over 8} (\dot{\varphi}_{n}+2\dot{\varphi}_{n+1}) } \rightarrow \infty$ the Laplacian $\Delta$ becomes
$$
\Delta= { 4n \over (8-n) } { \partial \over \partial \dot{\varphi}_{V} } { \partial \over \partial \dot{\varphi}_{V} } - \left( { 4n^{2} -2n  \over 8-n   } \right) { \partial \over \partial \dot{\varphi}_{V} } + { \partial \over \partial \dot{\varphi}_{g} } { \partial \over \partial \dot{\varphi}_{g} }  + { \partial \over \partial \dot{\varphi}_{g} } + \Delta_{SL(n)}
\eqno(3.6.3)
$$
where we have defined $\dot{\varphi}_{V}=\dot{\varphi}_{n} +2\dot{\varphi}_{n+1} $ and $\dot{\varphi}_{g}=-{3 \over 2}\dot{\varphi}_{n} + \dot{\varphi}_{n+1} $.
\par
The generic higher derivative term in the $d$ dimensional effective action was given in equation (3.1.3) and we  now examine how the automorphic form that it contains generically behaves in the large volume limit of the type IIA torus.  To proceed we will  convert the $d$-dimensional Planck length $l_{d}$ to the type IIA ten dimensional Planck length $l_{10 (A)}$ and the volume of the type IIA torus $V_{n(A)}$ using the relation 
$$
l_{d}=l_{10(A)} V_{n(A)}^{-{1 \over 8-n}}
\eqno(3.6.4)
$$
and use the condition
$$
\lim_{V_{n(A)} \rightarrow \infty} l_{10(A)}^{n} \int d^{d} x \sqrt{-g} V_{n(A)} = \int d^{10}x \sqrt{-\hat{g}}, 
\eqno(3.6.5)
$$
which implies that any term linear in $V_{n(A)}$ is preserved in the limit while any term with a lesser power of $V_{n(A)}$ vanishes in the limit.  In the $V_{n(A)} \rightarrow \infty$ limit $E_{n+1}$ decomposes as $GL(1) \times GL(1)  \times SL(n)$ and the $E_{n+1}$ coefficient function $\Phi_{E_{n+1}}$ generically splits into an expansion in the volume of the type IIA torus $V_{n(A)}$ and type IIA string coupling $g_{s(A)}$ as
$$
\Phi_{E_{n+1}}= \sum_{i} V_{n(A)}^{ a_{i}} g_{s(A)}^{c_{i}} \Phi_{SL(n)}^{i} 
\eqno(3.6.6)
$$
where $i$ labels the different $SL(n)$ coefficient functions arising in the limit and $a_{i}$ and $c_{i}$ are real numbers.   Demanding that the large volume limit of this generic higher derivative term converges to an acceptable higher derivative term in the type IIA effective action means that the large volume limit $V_{n(A)} \rightarrow \infty$ exists and that the resulting terms are ten dimensional type IIA higher derivative terms have constant coefficients   since the ten dimensional type IIB effective action can not depend on the moduli of the $n$ torus.  This condition may be expressed as
$$
l_{d}^{k-d}  \int d^{d}x \sqrt{-g}\Phi_{E_{n+1}} {\cal{O}} = l_{10 (A)}^{k-10}  \int d^{10}x \sqrt{-\hat{g}} \lim_{V_{n(A)} \rightarrow \infty} V_{n(A)}^{{2-k \over 8-n}} (\sum_{i} V_{n(A)}^{ a_{i}} g_{s(A)}^{c_{i}} \Phi_{SL(n)}^{i} ) {\cal{O}}
$$
$$
=l_{10 (A)}^{k-10}  \int d^{10}x \sqrt{-\hat{g}} (\sum_{i} b_{i} g_{s(A)}^{c_{i}}  ) \hat{{\cal{O}}}
\eqno(3.6.7)
$$
where $b_{i}$ is a real number and $\hat{\cal{O}}$ labels the different $d=10$ type IIA polynomials in the ten dimensional curvature $\hat{R}$, and field strengths $\hat{F}$ that arise in the decompactification of the $d$ dimensional polynomial in the curvature $R$, Cartan forms $P$ and field strengths $F$.  The last line of equation (3.6.7) expresses the above requirement that the only allowed coefficient functions in the ten dimensional type IIA theory are powers of $g_{s(A)}$ with trivial, that is, constant  $SL(n)$ automorphic forms.  The terms that are clearly preserved in this limit are those in $\sum_{i} V_{n(A)}^{ a_{i}}  g_{s(A)}^{c_{i}} \Phi_{SL(n)}^{i} $ with $V_{n(A)}^{-\left({2-k \over 8-n} \right)}$, in this case the factor of $V_{n(A)}$ combines with that contributed from converting the $d$ dimensional Planck length to the type IIA ten dimensional Planck length and $V_{n(A)}$ via equation (3.6.4) to converge to a ten dimensional type IIA higher derivative term.  Terms with a lesser power of $V_{n(A)}$ vanish in the $V_{n(A)} \rightarrow \infty$, while those with a greater power are non-analytic and must be treated carefully.
\par
In addition, the perturbative terms remaining after taking the limit must agree with a perturbative expansion in the ten dimensional type IIA string coupling $g_{s(A)}$.  In string frame this implies that each term has a $g_{s(A)}$ dependence of the form $g_{s(A)}^{-2 + 2g}$, where $g$ is the genus.  String frame in ten dimensions is related to Einstein frame by $g_{E \mu \nu}=e^{-{1 \over 2}\tilde{\sigma}}g_{S \mu \nu}$.  Upon rescaling to Einstein frame in the type IIA ten dimensional theory we find
$$
\int d^{10} x \sqrt{-g_{S}}  {\cal O}_{S}\to \int d^{10} x \sqrt{-g_{S}} g_{s(A)}^{ {\Delta-5 \over 2}} {\cal O}_{S},
\eqno(3.6.8)
$$
where ${\cal O}$ is some polynomial in the ten dimensional curvature $R$, fields strengths $F$ or derivatives of the type IIA dilaton, $S$ denotes string frame quantities and $\Delta$ is the number of ten dimensional type IIA space time metrics minus the number of inverse space time metrics.  Therefore any term that is preserved in the large volume limit of the type IIA torus must satisfy
$$
=l_{s(A)}^{k-10}  \int d^{10}x \sqrt{-\hat{g_{S}}} (\sum_{i} b_{i} g_{s(A)}^{c{i}}  )g_{s(A)}^{ {\Delta-5 \over 2}} \hat{{\cal{O}}_{S}} = l_{s(A)}^{k-10}  \int d^{10}x \sqrt{-\hat{g_{S}}} (\sum_{i} b_{i} g_{s(A)}^{-2+2g_{i}}  ) \hat{{\cal{O}}_{S}}
\eqno(3.6.9)
$$
where $g_{i}$ is the genus associated with the perturbative contribution to term $i$ in the large $V_{n(A)}$ expansion of the $E_{n+1}$ automorphic form.

\medskip
\noindent
{\bf {4. Derivation of Poisson equations}}
\bigskip
It has been found that demanding that the effective action  be invariant under supersymmetry implies that the automorphic form that appears as the coefficient of the $R^{4}$ term in the ten dimensional type IIB effective action satisfies a Laplace equation for which the  Laplacian is the one defined on the coset space of the massless scalar fields [9].  The corresponding Laplace, or Poisson equations,  satisfied by the automorphic forms that appear as the coefficient functions in the effective action for the higher order terms and in $d<10$ dimensions have not been deduced directly via supersymmetry constraints.  However, the automorphic forms that occur for  the $R^{4}$, $\partial^{4} R^{4}$ and $\partial^{6} R^{4}$ terms in $d= 10-n$ dimensions have been conjectured and found to lead to all the known and perturbative and non-perturbative features of these terms [20,21,22] and these are also known to satisfy   the Poisson equations.   In particular, the automorphic forms that occur as coefficients of $R^{4}$, $\partial^{4} R^{4}$ in $d= 10-n$ dimensions, denoted 
$\Phi_{E_{n+1}}^{R^4}$ and $\Phi_{E_{n+1}}^{\partial^{4} R^4}$ respectively, are expected to satisfy the equations [20,21,22]
$$
\Delta_{E_{n+1}} \Phi_{E_{n+1}}^{R^4} + {3(n^{2}-n-2) \over 8-n}  \Phi_{E_{n+1}}^{R^4} = 0
\eqno(4.1)
$$ 
and 
$$
\Delta_{E_{n+1}} \Phi_{E_{n+1}}^{\partial^{4} R^4} + {5(n^{2}-n-6) \over 8-n}   \Phi_{E_{n+1}}^{\partial^{4} R^4} = 0
\eqno(4.2)$$
where where $\Delta_{E_{n+1}}$ is the $E_{n+1}$ Laplacian, which is given in appendix A.
\par
However the automorphic form $\Phi_{E_{n+1}}^{\partial^{6} R^4}$ which is the coefficient of $\partial^{6} R^{4}$
obeys a more complicated equation namely [20,21,22]
$$
\Delta_{E_{n+1}} \Phi_{E_{n+1}}^{\partial^{6}R^{4}} + {6(n^{2}-16) \over 8-n}  \Phi_{E_{n+1}}^{\partial^{6}R^{4}} = - \left( \Phi_{E_{n+1}}^{R^{4}} \right)^{2}.
\eqno(4.3)
$$
 As we mentioned in the introduction if one knew the automorphic forms that occur in the effective action then one would know all 
string effects, at least for ten dimensions and for toroidal compactifications. For certain dimensions the above Poisson equations contain constants on the right-hand, these are connected to non-analytic terms that we do not consider in this paper. 
\par
In this section we give a different approach to the problem of determining the automorphic forms that is based on the behaviour of the Laplacian and the automorphic form in the limit when one of the  dimensions is decompactified. Our aim is to use this limit to place restrictions on the differential equation that the automorphic form can obey. This is particularly useful as knowing the differential equation one can using the formulae given in this paper to deduce the behaviour of the automorphic form 
in all the possible different limits; indeed knowing the equation is almost 
tantamount to knowing  the automorphic form itself.  We now give two assumptions that place very strong restrictions on the coefficients that occur in the Poisson equation. 
\medskip
\noindent
{\bf Assumption 1}

We assume that the automorphic form obeys a differential equation of the form

$$
\Delta_{E_{n+1}} \Phi_{E_{n+1}}^{R^{{k \over 2}}} + A^k(n) \Phi_{E_{n+1}}^{R^{{k \over 2}}} = \sum \prod_{i}B_{i}^k(n) ({\Phi}_{E_{n+1}}^{R^{{k_i \over 2}}})^{a_{i}^k}
\eqno(4.4)
$$
where $A^k(n)$, $B_{i}^k(n)$ are constants and $a_{i}^k(n)$ are integers, the sum on the right hand side is over all possible products of coefficient functions appearing at lower orders in the effective action than $\Phi_{E_{n+1}}$.   
Clearly this assumption is true for the cases when the number of space-time derivatives in the effective action is fourteen or less. As the Laplacian acting on an automorphic form is also an automorphic form,  the right hand side must also be an automorphic form and so this assumption really amounts to the assumption that the  automorphic form which occurs on the right hand side of the equation is composed of the automorphic forms that occurred for lower number of space-time derivatives. One might suspect that this can be shown in general from the supersymmetric nature of the effective action. 
\par
As mentioned in section 3.4, in the decompactification of a single dimension limit we take $ \left( { r_{d+1} \over l_{d+1}} \right) \rightarrow \infty$ and the higher derivative terms in the $d$ dimensional effective action lift to higher derivative terms in the $d+1$ dimensional effective action.  In this limit the $E_{n+1}$ automorphic  functions of the higher derivative terms in the effective action decompose into $GL(1) \times E_{n}$ automorphic forms where the $GL(1)$ factor is associated with the  power of ${ r_{d+1} \over l_{d+1}}$.  
\medskip
\noindent
{\bf Assumption A2}

The automorphic form  $\Phi^{R^{{k \over 2}}}_{E_{n+1}}$ associated with  a higher derivative term in the $d=10-n$ dimensional effective action, where $k$ denotes the number of $d$ dimensional space-time  derivatives in the higher derivative term, decompactifies as 
$$
\lim_{ \left( {r_{d+1} \over l_{d+1}} \right) \rightarrow \infty} l_{d}^{k-d} \Phi^{R^{{k \over 2}}}_{E_{n+1}}= l_{d+1}^{k-d}  \{ c^{{k\over 2}}_{{k\over 2}} \left( {r_{d+1} \over l_{d+1}} \right) \Phi^{R^{{k \over 2}}}_{E_{n}} + \sum_{j<k} c^{{k\over 2}}_{{j\over 2}} \left( {r_{d+1} \over l_{d+1}} \right)^{2+k-j-d} \Phi^{R^{{j \over 2}}}_{E_{n}} 
$$
$$+  c_{{k\over 2}} ^{0} \left( {r_{d+1} \over l_{d+1}}  \right)^{k-d} 
 +\dots \} .
\eqno(4.5)
$$
where $c^{{k\over 2}}_{{j\over 2}}$ are constants and  the sum is over all coefficient functions $\Phi^{j}_{E_{n}}$ of higher derivative terms satisfying $j<k$ where $j$ is the number of derivatives of the associated term. The $+\ldots$ denoted certain terms that are required for consistency and arise from using the decompactification limit on the terms that occur on the right hand side of equation (4.4); these terms are known and may be derived by an induction procedure from the terms that have been explicitly written down in equation (4.5). We will later on give an example of how this works. This expansion is consistent with those given in [20,21,22] for the cases of $R^4$, $\partial^4R^4$ and $\partial^4R^4$.  
The $l_{d}^{k-d}$ factor multiplying $\Phi^{k}_{E_{n+1}}$ arises since the higher derivative terms in the effective action are of the form 
$$
l_{d}^{k-d} \int d^{d} x \sqrt{-g} \Phi^{ {\cal O}}_{E_{n+1}} {\cal O}
$$
where ${\cal O}$ is a $k$ derivative polynomial in the $d$ dimensional type II string theory, curvatures $R$, field strengths $F$ and Cartan forms $P$.
Using equation (3.4.4) we find this expansion can be written in the form 
$$
\lim_{ \left( {r_{d+1} \over l_{d+1}} \right) \rightarrow \infty}  \Phi^{k}_{E_{n+1}}=  ({r_{d+1} \over l_{d+1}})^{{k-d\over 8-n}} \{ c^{{k\over 2}}_{{k\over 2}} \left( {r_{d+1} \over l_{d+1}} \right) \Phi^{k}_{E_{n}} + \sum_{j<k} c^{{k\over 2}}_{{j\over 2}} \left( {r_{d+1} \over l_{d+1}} \right)^{2+k-j-d} \Phi^{j}_{E_{n}} 
$$
$$
+  c_{0}^{{k\over 2}}\left( {r_{d+1} \over l_{d+1}}  \right)^{k-d} 
+ \ldots \}.
\eqno(4.6)
$$
\par
When we substitute the decompactification limit of the automorphic form of equation (4.6) into equation (4.4) we find a set of equations,  one equation for each power of ${r_{d+1} \over l_{d+1}} $ that occurs. As we will see these place very strong conditions on the coefficients that occur in  equation (4.4). 
\par
The behaviour of the first term in equation (4.5), or (4.6), is determined by demanding that one finds in the decompactified effective action a term  of the form 
$$
l_{d+1}^{k-d-1}\int d^{d+1} x \sqrt {-g}\Phi^{k}_{E_{n}}R^{{k\over 2}}
\eqno(4.7)$$
Indeed looking at equation (4.6) we see that the factor of ${r_{d+1} \over l_{d+1}} $ is required to reproduce  the measure and   the change from the factor of the Planck length  $l_d$ to $l_{d+1}$ which is required to get the appropriate dimensional factor in $d+1$ dimensions. Thus the first terms in equation (4.5), or (4.6),  is not an assumption but can be shown to be true using the above argument.

Comparing the leading order power, that is $ {r_{d+1} \over l_{d+1}} ^{{k-d\over 8-n}+1}$, which occurs on the left-hand side of equation  (4.5), or (4.6), that is in $\Delta_{E_{n+1}} \Phi_{E_{n}} + A(n) \Phi_{E_{n}} $,  with that which arises from the powers of the automorphic forms on the right-hand side we find that  
$$
{k-d\over 8-n}+1=\sum_i ({k_i-d\over 8-n}+1)a_{i}^{k}
\eqno(4.8)$$
and so resulting in the condition 
$$
k=\sum_i (k_i-2)a_{i}^{k}+2
\eqno(4.9)$$
This places a strong constraint on the automorphic forms that can occur on the right-hand side of equation (4.4). We note that we have deduced  the first power of ${r_{d+1} \over l_{d+1}} $ on the right-hand side of (4.5), or (4.6),   without making any assumption and so the results just derived have a similar status. 
\par
We can also compare the coefficient of this leading term. Using equation (2.7) we find that the leading power can be written as ${r_{d+1} \over l_{d+1}} ^{{k-d \over 8-n}+1}= e^{{(k-2) \dot{\varphi}_{1} \over 9-n}}$. Substituting into equation (4.4), only keeping terms with this power, using the Laplacian in the decompactification limit of equation (3.4.2) we find that the coefficients of equation (4.4) must obey 
$$
A^k(n) -A^k (n-1) = -{(k-2)(k+10+n^2-17n)\over 2 (8-n)(9-n) }
\eqno(4.10)
$$
The eigenvalues $A^k(n)$ were deduced in reference [20] for $k=8, 12$ and $14$ using the leading power in the expansion of the automorphic form in
the decompactification limit, as we have just done, but then using its known value in ten dimensions.  In this paper we also find the eigenvalues from equation (4.10) but using additional equations which result from the behaviour of the Poisson equation under the other terms in the expansion of the automorphic form in the limit. 

\par
We will now use  assumptions A1 and A2 to place further conditions on the coefficients that occur in equation (4.4) by comparing the coefficients of the other powers of ${r_{d+1} \over l_{d+1}} $. 
To illustrate how this works in a simple way we will  assume  that there are no $+\ldots$ contributions in equation (4.5), or (4.6), that is, there are  no contribution from the automorphic forms on the right hand side of equation (4.4) that lead to   powers of ${r_{d+1} \over l_{d+1}} $ in these equations.  This is not always the case and if there is then the equations (4.12) and (4.14) below must be modified accordingly. However, making this assumption will allow us to demonstrate the power of the method.  Comparing the coefficients of 
$$
({r_{d+1} \over l_{d+1}}) ^{{(d-k)(n-9)\over (8-n)}}= e^{{(k-d)}\dot \varphi_{1}}
\eqno(4.11)$$
we find  terms which contain  no automorphic forms and which lead to the condition 
$$
A^k(n)=-{(k-10+n)(9k-78-n(k-2))\over 2(8-n)}
\eqno(4.12)$$ 
We note that this does indeed satisfy equation (4.10).
\par
Finally let us compare the remaining coefficients, that is,   those that occur with the  powers 
$$
({r_{d+1} \over l_{d+1}}) ^{k-k_i +2 -d + {(k-d)\over (8-n)}}
= e^{{((8-n)(k-k_i +n-8)+k-10+n)\over (9-n)}\dot \varphi_{1}}
\eqno(4.13)$$
Using the Poisson equation for $\Phi^{k_i}$ in $d+1$ dimensions we find the  constraints 
$$
A^k(n) -A^{k_i}(n-1) = -{(k(9-n)+k_i(n-8) -62)((8-n)(k-k_i +n-8)+k-10+n)\over 2 (8-n)(9-n) }
\eqno(4.14)
$$
As we have mentioned if there are $+\ldots $ contributions in equation (4.5), or equation (4.6) the above results are modified. Below we show how this works for the case of $\partial^6 R^4$. 
\par
We now show that assumptions A1 and A2 lead to the known equations satisfied by the automorphic form associated with the $R^{4}$, $\partial^{4} R^{4}$ and $\partial^{6} R^{4}$ terms in the $d=10-n$ dimensional type II string effective action. Let us begin with the $R^{4}$. For this case equation (4.6)  reads 
$$
\lim_{ \left( {r_{d+1} \over l_{d+1}} \right) \rightarrow \infty}  \Phi^{R^{4}}_{E_{n+1}} = ({r_{d+1} \over l_{d+1}})^{{8-d}\over 8-n} \left( c^4_4 \left( {r_{d+1} \over l_{d+1}} \right) \Phi^{R^{4}}_{E_{n}} + c^4_0 \left( {r_{d+1} \over l_{d+1}} \right)^{8-d}  \right),
\eqno(4.15)
$$
Since there are no automorphic forms corresponding to terms with fewer space-time derivatives,  equation (4.9) implies that the equation $\Phi^{R^{4}}_{E_{n+1}}$ satisfies has no right-hand side and so is of the form 
$$
\Delta_{E_{n+1}} \Phi_{E_{n+1}}^{R^{4}} + A(n)^{R^{4}} \Phi_{E_{n+1}}^{R^{4}} = 0,
\eqno(4.16)
$$
\par
In the $ \left( { r_{d+1} \over l_{d+1}} \right) \rightarrow \infty$ limit equation (4.16) becomes
$$
\left( { (9-n) \over 2(8-n) } { \partial \over \partial \dot{\varphi}_{1} } { \partial \over \partial \dot{\varphi}_{1} } - { (-n^{2} + 17n - 12) \over 2(8-n)   } { \partial \over \partial \dot{\varphi}_{1} } + \Delta_{E_{n}} + A(n)^{R^{4}} \right) 
$$
$$\times\left( c^{4}_{4}e^{{6 \over 9-n}\dot{\varphi}_{1}} \Phi^{R^{4}}_{E_{n}} +  c^{4}_{0}e^{{(-2+n) }\dot{\varphi}_{1}} \right) = 0
\eqno(4.17)$$
where we have used the expression of equation (3.4.2) for the Laplacian in this limit and we have used equation (2.7) to write $\left( { r_{d+1} \over l_{d+1}} \right)$ in terms of $\dot{\varphi}_{1}$. Collecting terms that contain 
$e^{{(-2+n)\dot \varphi_1}} $ we find that 
$$
A(n)^{R^{4}} = {3(n^{2}-n-2) \over 8-n}
\eqno(4.18)
$$
This agrees with the known value as given in equation (4.1). The only other powers of  $\left( { r_{d+1} \over l_{d+1}} \right)$ are given by  $e^{{6 \over 9-n}\dot{\varphi}_{1}} $ and these imply that 
$$
A_{n}^{R^{4}} - A_{n-1}^{R^{4}} = {3 (-n^2 +17n -18)\over (9-n) (8-n) }
\eqno(4.19)$$
This is indeed satisfied by the above values of $A_{n}^{R^{4}}$ and agrees with equation (4.14) when $k=8$. Thus we have recovered from assumptions A1 and A2 the equation satisfied by the automorphic form which is the coefficient in the effective action of $R^4$. 
\par
We now consider the automorphic form $ \Phi_{E_{n+1}}^{\partial^{4} R^{4}}$ that is the coefficient of the $\partial^{4} R^{4}$. Examining equation (4.9) we find that this equation also possess no right-hand side and so has the form 
$$
\Delta_{E_{n+1}} \Phi_{E_{n+1}}^{\partial^4R^{4}} + A(n)^{\partial^4R^{4}} \Phi_{E_{n+1}}^{\partial^4 R^{4}} = 0,
\eqno(4.20)
$$
For this automorphic form the decompactification limit of equation (4.5) 
reads
$$
\lim_{ \left( {r_{d+1} \over l_{d+1}} \right) \rightarrow \infty}  \Phi^{\partial^{4} R^{4}}_{E_{n+1}} 
$$
$$
= ({r_{d+1} \over l_{d+1}})^{{12-d}\over 8-n} \left( c^6_6 \left( {r_{d+1} \over l_{d+1}} \right) \Phi^{\partial^{4} R^{4}}_{E_{n}} +  c^6_4\left( {r_{d+1} \over l_{d+1}} \right)^{6-d} \Phi^{R^{4}}_{E_{n}} +  c^6_0\left( {r_{d+1} \over l_{d+1}}  \right)^{12-d} \right),
\eqno(4.21)
$$
Using equation (2.7) and (3.4.2) this equation is given in the decompactification limit by 
$$
\left( { (9-n) \over 2(8-n) } { \partial \over \partial \dot{\varphi}_{1} } { \partial \over \partial \dot{\varphi}_{1} } - { (-n^{2} + 17n - 12) \over 2(8-n)   } { \partial \over \partial \dot{\varphi}_{1} } + \Delta_{E_{n}} + A(n)^{\partial^{4} R^{4}} \right) 
$$
$$
\times \left( c^{6}_{6}e^{{10 \over 9-n} \dot{\varphi}_{1}} \Phi^{\partial^{4} R^{4}}_{E_{n}} +  c^{6}_{4}e^{-{(10-n)(3-n) \over 9-n}\dot{\varphi}_{1}} \Phi^{R^{4}}_{E_{n}} +  c^{6}_{0}e^{{(2+n)}\dot{\varphi}_{1}} \right)  = 0,
\eqno(4.22)
$$
We can now compare the coefficients of the three powers that occur giving three equations. We have already analysed the leading power, that is $e^{{10 \over 9-n} \dot{\varphi}_{1}} $, and we find that for $k=12$ equation (4.14) becomes  
$$
A(n)^{ R^{4}} - A(n-1)^{\partial^{4} R^{4}} = -{(2n+9)(n-3)(n-10)\over (8-n)(9-n) }
\eqno(4.23)$$
The coefficients of the power $e^{-{(10-n)(3-n) \over 9-n}\dot{\varphi}_{1}} $
implies the equation 
$$
A(n)^{\partial^{4} R^{4}} - A(n-1)^{\partial^{4} R^{4}}= -5{(22+n^2-17n)\over (8-n)(9-n) }
\eqno(4.24)$$
Finally equating to zero the terms that occur with the power $e^{{(2+n)}\dot{\varphi}_{1}} $ we find that 
$$
A(n)^{\partial^{4} R^{4}} = {5(n^{2}-n-6) \over 8-n}.
\eqno(4.25)
$$
With this value and that of equation (4.14) we find that equations (4.23)
and (4.24) are automatically satisfied. 
\par
We now repeat the procedure for the coefficient of $\partial^{6} R^{4}$, that is, the automorphic form  denoted by $\Phi^{\partial^{6} R^{4}}_{E_{n+1}} $.  Assuming the Poisson equation satisfied by $\Phi^{\partial^{6} R^{4}}_{E_{n+1}} $ is of the form (4.4) we find in this case that the right hand side of the Poisson equation can be non-zero.  In particular, one finds that the condition of equation (4.9) for the possible polynomials of the automorphic forms found at lower orders in the effective action does have the  solution $i=8$ and $a_{i}=2$ for $k=14$ and so one can have  on the right hand side of the Poisson equation  the term
$$
B(n) (\Phi^{R^{4}}_{E_{n+1}})^{2}.
\eqno(4.26)
$$
As a result the Poisson equation satisfied by $\Phi^{\partial^{6} R^{4}}_{E_{n+1}} $ is given by 
$$
\Delta_{E_{n+1}} \Phi_{E_{n+1}}^{\partial^{6}R^{4}} + A(n)^{\partial^{6}R^{4}} \Phi_{E_{n+1}}^{\partial^{6}R^{4}} =  B (\Phi^{R^{4}}_{E_{n+1}})^{2}.
\eqno(4.27)
$$
Expanding the right hand side of equation (4.23) using equation (4.6) we have
$$
\lim_{{r_{d+1} \over l_{d+1}} \rightarrow \infty}  B(n) (\Phi^{R^{4}}_{E_{n+1}})^{2} = B(n) \{(c^{4}_{4})^{2} {r_{d+1} \over l_{d+1}}^{12 \over 8-n} (\Phi_{E_{n+1}}^{R^{4}})^{2} 
$$
$$
+ 2 c^{4}_{4} c^{4}_{0} \left({r_{d+1} \over l_{d+1}}\right)^{-n^{2} +11n -12 \over 8-n} \Phi_{E_{n+1}}^{R^{4}} 
+ (c^{4}_{0})^{2} \left({r_{d+1} \over l_{d+1}}\right)^{2 (9-n)(n-2) \over 8-n} \}
\eqno(4.28)
$$
We note that there this expression contains the automorphic form $\Phi_{E_{n+1}}^{R^{4}}$ squared in $d+1$  dimensions but also the automorphic form $\Phi_{E_{n+1}}^{R^{4}}$ which is required for the Poisson equation in $d+1$ dimensions. It also contains a term with no automorphic form which will lead to such a term on the right-had side of the Poisson equation. 
\par
Using equation (4.5) the   expansion  of $\Phi^{\partial^{6} R^{4}}_{E_{n+1}} $ in the decompactification limit is given by 
$$
\lim_{{r_{d+1} \over l_{d+1}} \rightarrow \infty} l_{d}^{14-d} \Phi^{\partial^{6} R^{4}}_{E_{n+1}} 
$$
$$
= l_{d+1}^{14-d} \left( c^{7}_{7} \left( {r_{d+1} \over l_{d+1}} \right) \Phi^{\partial^{6} R^{4}}_{E_{n}} +  c^{7}_{6} \left( {r_{d+1} \over l_{d+1}} \right)^{4-d} \Phi^{\partial^{4} R^{4}}_{E_{n}} + c^{7}_{4} \left( {r_{d+1} \over l_{d+1}} \right)^{8-d} \Phi^{R^{4}}_{E_{n}} \right.
$$
$$
+c^{7}_{0} \left. \left( {r_{d+1} \over l_{d+1}} \right)^{14-d} + d^{7}_{0} \left( {r_{d+1} \over l_{d+1}} \right)^{15-2d}  \right).
\eqno(4.29)
$$
In this expansion we find two terms which contain no automorphic form. The first of which is the one expected and listed explicitly in equation (4.5), while the second term is one of those whose presence was indicted by the $+\ldots$ and arises to compensate such a term that appears on the right-hand side; indeed the final term in equation (4.28). 
\par
Using equations (4.4), (4.6) and (4.23) the Poisson equation in the decompactification limit is given by
$$
\left( { (9-n) \over 2(8-n) } { \partial \over \partial \dot{\varphi}_{1} } { \partial \over \partial \dot{\varphi}_{1} } - { (-n^{2} + 17n - 12) \over 2(8-n)   } { \partial \over \partial \dot{\varphi}_{1} } + \Delta_{E_{n}} + A(n)^{\partial^{6} R^{4}} \right) 
$$
$$
\times { c^{7}_{7} e^{{12 \over 9-n} \dot{\varphi}_{1}} \Phi^{\partial^{6} R^{4}}_{E_{n}} + c^{7}_{6} e^{-(11-n)(4-n) \over 9-n} \dot{\varphi}_{1}} \Phi^{\partial^{4} R^{4}}_{E_{n}} + c^{7}_{4} e^{{-n^{2}+11n-12 \over 9-n} \dot{\varphi}_{1}} \Phi^{ R^{4}}_{E_{n}} 
$$
$$
+ c^{7}_{0} e^{(4+n) \dot{\varphi}_{1}} + d^{7}_{0} e^{(2n-4) \dot{\varphi}_{1}} \}
$$
$$
= B(n)((c^{4}_{4})^{2} e^{{12 \over 9-n} \dot{\varphi}_{1}} (\Phi_{E_{n+1}}^{R^{4}})^{2} + 2 c^{4}_{4} c^{4}_{0} e^{{-n^{2} +11n -12 \over 9-n} \dot{\varphi}_{1}} \Phi_{E_{n+1}}^{R^{4}} + (c^{4}_{0})^{2} e^{{2(-2+n)} \dot{\varphi}_{1}} ).
\eqno(4.30) 
$$ 
Equating the coefficient of $e^{(4+n) \dot{\varphi}_{1}}$ we find that 
$$
A(n)^{\partial^{6} R^{4}} = {6(n+4)(n-4) \over 8-n}.
\eqno(4.31)
$$
in agreement with the known values. The coefficients of the terms containing $e^{{12 \over 9-n} \dot{\varphi}_{1}}$ imply that 
$$
A(n)^{\partial^{6} R^{4}}- A(n-1)^{\partial^{6} R^{4}}= -{ 6(n^2-17n+24) \over (8-n)(9-n)}, 
\eqno(4.32)
$$
and, examining the Poisson equation that results in $d+1$ dimensions,  that 
$$
(c^4_4)^2 B(n)= B(n-1)
\eqno(4.33)
$$
the coefficients of the terms containing $e^{-{(11-n)(4-n) \over 9-n} \dot{\varphi}_{1}}$ imply that 
$$
A(n)^{\partial^{6} R^{4}}- A(n-1)^{\partial^{4} R^{4}}= -{(n-11)(n-4)(n+16) \over (8-n)(9-n)},
\eqno(4.34)
$$
the coefficients of the terms containing $e^{{-(n^2-11n+12) \over 9-n} \dot{\varphi}_{1}}$ imply that 
$$
A(n)^{\partial^{6} R^{4}}- A(n-1)^{R^{4}}=  2 B(n) { c^{4}_{0} c^{4}_{4} \over c^{7}_{4} } - {3n(n^{2}-11n+12) \over (8-n)(9-n)} 
\eqno(4.35)
$$
and finally the coefficients of the terms containing $e^{2(n-2) \dot{\varphi}_{1}}$ imply the equation
$$
A(n)^{\partial^{6} R^{4}}=B(n) { (c^{4}_{0})^{2} \over d^{7}_{0}} - {(n-2)(-n^2 +5n-24)\over (8-n)}.
\eqno(4.36)
$$
\par 
Using equation (4.31) and equation (4.19) we may write equation (4.35) as 
$$ 
B(n) {c_0^4 c^4_4 \over c_4^7}=-3{(18n^2 +44n +288) \over (9-n)(8-n)}
\eqno(4.37)
$$
While using equation (4.31) we may write equation (4.36) as 
$$
(n-6)(n+1) {d_0^7\over (c_0^4)^2}= B(n)
\eqno(4.38)$$
\par
For all the automorphic forms we have considered the pattern is the same, the eigenvalue for $n$ is determined by the terms with no automorphic form in the decompactification and then the  eigenvalues for other values of $n$ by the terms that give the same automorphic form back, for example in equations (4.31) and (4.32) respectively. 
\par
By normalising the way the automorphic form occurs in the effective action we may choose the coefficient $c_k^k=1$. Equation (4.33) then implies that $B(n) \equiv B$ is independent of $n$. With these values equations (4.37) and (4.38) simplify and place strong constraints on the coefficient functions. 
\par
In carrying out the above calculations we have used independent decompactification formulae for the automorphic forms involved, however, the  powers of $e^{ \dot{\varphi}_{1}}$ coming from the right hand-side from the decompactification of 
$\Phi^{ R^{4}}$ must match those coming from the decompactification of $\Phi^{\partial^{6}} R^{4}$. This places a strong check on the decompactification formulae we have used.

\par
We close this section by giving an alternative derivation of equation (4.9). 
By considering the dimensional reduction of the effective action from eleven dimensions it was argued [29] that the automorphic form $\Phi ^{R^{{k\over 2}}}$ that occurs in the term in the effective action of the form 
$\int d^d x \Phi ^{R^{{k\over 2}}}R^{{k\over 2}}$
had to contain a term containing the exponential 
$$
\Phi ^{R^{{k\over 2}}}\sim e^{-\sqrt 2 ({k-2\over 6}) \underline \varphi \cdot \underline \Lambda _{n+1}}
\eqno(4.39)$$
where $\underline \Lambda _{n+1}$ is the fundamental weight of $E_{n+1}$ associated with node $n+1$ and as before $k$ is the number of space-time derivatives in the  automorphic form. As explained in reference [27] this is consistent with what is  known about the automorphic forms that are known to occur. 
This exponential factor arises from the automorphic form on the left-hand side of equation (4.4) and as a result it must also occur on the right-hand side of this equation if this contains a product  automorphic forms.  As a result the  sum of the weights from the automorphic forms on the right  hand side of Poisson equation must match those from the automorphic form on the left-hand side and so we immediately find the condition of equation (4.9).   One can find the same conclusion by using the constraints that arise form the dimensional reduction of the IIA [28] or IIB theories [27].


\medskip 
\noindent
{\bf 5. The Poisson equation and its perturbative
 limit} 
\medskip 

\par
In this section we will consider how the Poisson equation (4.4) behaves in the perturbative limit studied in section three. There the perturbative limit appears as taking $g_d= e^{-{(8-n) \over 4}\dot{\varphi}_n}\to 0$. This breaks $E_{n+1}$ into $SO(n,n)\times GL(1)$. For simplicity we will restrict our attention to the ten-dimensional IIB theory in which case $n=0$ and so $g_d= e^{-2\dot{\varphi}_0}$ and we have an $SL(2,R)$ symmetry, but one could carry out the analysis for the IIA theory and indeed in any dimension. The string coupling is given in terms of the dilaton $\phi$ of the IIB by $g_s= e^\phi$ and so $\phi =-2\dot{\varphi}_0$. In the perturbative limit the Laplacian of equation (3.2) is given by 
$$
\Delta = {d^2\over d\phi^2} + {d\over d\phi}
\eqno(5.1)$$
\par
 Let us consider a  contribution to the effective action describing graviton scattering which can be written in the generic form  
$$
\int d^{10} x \sqrt{-g_{E}} \Phi_{SL(2)}^{{k\over 2}} R^{{k\over 2}}_{E}
\eqno(5.2)
$$
where the subscript $E$  denotes that we are in Einstein frame. 
Converting to string frame through the redefinition $(g_{E})_{\mu \nu}=e^{-{1 \over 2} \phi} (g_{S})_{\mu \nu}$, where $\phi$ is the dilaton and $S$ subscript denotes string frame quantities, one has
$$
\int d^{10} x \sqrt{-g_{E}}R^{m}_{E} = \int d^{10} x \sqrt{-g_{S}}  \Phi_{SL(2)} e^{- \left({5-m \over 2}\right) \phi }R_{S}^{m}.
\eqno(5.3)
$$
where $m={k\over 2}$. 
In the perturbative limit, that is  $g_{s} = e^{\phi} \rightarrow 0$,  we require that in string frame a $k$ derivative term in the $d=10$ dimensional effective action constructed from $m$ curvatures takes the form
$$
\int d^{10} x \sqrt{-g_{S}} \sum_{q=0}^{\infty} a_{q} e^{\left( -2 + 2q \right) \phi }R_{S}^{m}.
\eqno(5.4)
$$
where $q$ is the genus, or loop order, of the corresponding perturbative contribution and $a_{q}$ are the associated real coefficients.  Therefore the perturbative contribution from the coefficient function $\Phi_{SL(2)}$ must be of the form
$$
\lim_{\phi \rightarrow 0} \Phi_{SL(2)}= \sum_{q=0}a_{q}e^{ \left( {5-m \over 2} -2 + 2q  \right) \phi}.
\eqno(5.5) 
$$
\par
The perturbative contribution to the automorphic form can be thought of as composed of a homogeneous solution, for which the right hand side vanishes, and a particular solution, both of which are a power series in the string coupling $g_s$, in other words exponentials of the form  
$e^{-s\phi}$. Let us suppose that the automorphic form has a part which is a homogeneous solution, then equation (4.4) implies that the eigenvalue has the form 
$$
A = -s(s-1)
\eqno(5.6)$$
where $A\equiv A^{R^m}(0)$ is the coefficient that occurs in equation (4.4) in ten dimensions. 
\par
The automorphic forms for SL(2,R) that obey equation (4.4) with no right-hand side are well known and in the perturbative  limit they have the generic form 
$$
e^{-s\phi}+ e^{(s-1)\phi}
\eqno(5.7)$$
Comparing these with equation (5.5) we conclude that 
$$
s+ {5\over 2} -{m\over 2} = 2-2q_1 \quad {\rm and }\quad s- {7\over 2} +{m\over 2} = -2+2q_2
\eqno(5.8)$$ 
where $q_1$ and $q_2$ are positive integers which correspond to the orders of perturbation theory which occur in the homogeneous solution. 
Adding and subtracting these equations we find that 
$$
m=2(q_1+q_2+1) \quad {\rm and }\quad s=q_2-q_1+{1\over 2} 
\eqno(5.9)$$ 
Examining equation (5.8) we find the limits 
$$
s+{m\over 2}\ge {3\over 2}  \quad {\rm and }\quad {m\over 2}-s\ge {1\over 2}  
\eqno(5.10)$$
which in turn implies that $s(s-1)\ge {(m-1)(m-3)\over 4}$ and  $s(s-1)\le {
(m-1)(m-3)\over 4}$ from which we conclude that 
$$s(s-1)=  {(m-1)(m-3)\over 4} \quad {\rm and \ so} \quad s={m-1\over 2}
\eqno(5.11)$$
It follows that the resulting string contributions which arise from a homogeneous term to the Poisson equation are,  in string frame, of the generic form 
$$
e^{-2\phi} + e^{(m-4)\phi} ,
\eqno(5.12)$$
that is,  a tree and ${m-4 \over 2}$ loop contribution. 
\par
To see how  this works let us apply it to the much studied cases of $R^4$ and $R^6\sim \partial^4 R^4$ which have $m=4$ and $m=6$. The automorphic forms for these two cases obey equations (4.1) and (4.2) respectively both of which have no right hand side and so the 
homogeneous solution is the only solution. For $R^4$ we find that $s={3\over 2}$ and we have a tree level and one loop contribution while for 
$\partial^4 R^4$ we find that $s={5\over 2}$ and we have a tree level and two loop contribution. 
\par
Clearly if $m$ is an odd integer, for example for $\partial^6 R^4\sim R^7$, then the homogeneous solution is not compatible with string perturbation theory and we must conclude that the homogeneous solution is not present in the automorphic form. 
\par
We now consider the particular solution which by definition receives contributions from the right hand side. The factor that converts the automorphic form from Einstein frame to string frame  is given in equation (5.3) and this  can be written using equation (4.9) as 
$$
e^{-\phi {5-m \over 2} }= e^{-2\phi  } \Pi_i e^{\phi {(m_i -1) \over 2} }  
\eqno(5.13)
$$
To convert the Poisson equation to string frame we must multiply by this factor on the left and  right hand side of this equation. When doing this on the right hand side this is equivalent to  multiplying the individual automorphic forms that occur in the product by 
$$
\Phi ^{{k_i\over 2}} \to e^{\phi {(m_i -1) \over 2}}\Phi ^{{k_i\over 2}}  
\eqno(5.14)$$
where $2m_i=k_i$  and then multiply the result also  by $e^{-2\phi  }$. 
\par
This last maneuver makes it easy find the perturbative terms that are part of the particular solution and arise from the right hand side. Let us see how this works for the correction with the smallest number of space-time derivatives that has a right hand side, namely, the automorphic form that appears with  $\partial^6 R^4$ with $m=7$, or  $k=14$. This automorphic form obeys equation (4.3).  To transform this equation to string frame we must multiply the left hand side by
$e^{\phi}$.  On the right hand side we find the square of $\Phi^{R^4}$ and to get to string frame we multiply each of these factors by  $e^{{3\over 2}\phi}$, namely 
$$
e^{{3\over 2}\phi}\Phi^{R^4}\sim e^{0\phi}+ e^{{ 2}\phi}
\eqno(5.15)$$
As a result on the right hand side we find in the perturbative limit the terms 
$$
e^{{- 2}\phi} (e^{0\phi}+ e^{{ 2}\phi})^2= e^{{- 2}\phi}+ e^{{0}\phi}+ e^{{2}\phi}
\eqno(5.16)$$
Thus the $\partial^6 R^4$ term in  the effective action must have contributions to at tree level, one loop and two loop. 
\par
It is instructive  to continue with this example. The value of $A$  in this case follows by putting $n=0$ in equation (4.12)  to find that $A=-12$. We now consider the homogeneous solution to the Laplacian {\bf in the perturbative limit}. The solution will be of the form of equation (5.7) in this limit and so we take $12= s(s-1)$ and so $s=4$ or $-3$, for either choice the homogeneous solution  {\bf in the perturbative limit }
is given by  $e^{{- 4}\phi}+ e^{{3}\phi}  $. To transform to string frame we must, as noted above, multiply by $e^{\phi}$  to find 
$e^{{- 3}\phi}+ e^{{4}\phi} $. The first   term is not allowed in string perturbation theory but the second term is and this can then appear in the particular solution in addition to the terms that must appear as they appear on the right hand side. At first sight this is a contradiction as there is no homogeneous solution to the full Poisson equation in this case. However, 
it is important to distinguish between the homogeneous solution to the full Poisson equation and a homogeneous solution in the Poisson equation in the perturbative limit. We note that by definition,  in the perturbative limit,   a term in the homogeneous  solution does not appear in the right-hand side of the Poisson equation. The term in the homogeneous solution in the perturbative limit that is an acceptable string correction is in fact part of the particular solution to the full Poisson equation. 
Consequently the  $\partial^6 R^4$ contribution can  also have a three loop contribution in addition to the tree level, one loop and two loop contributions found above.   
\par 
One can iteratively repeat this procedure order by order for automorphic forms with increasing number of space-time derivatives in the level to find which   orders in perturbation theory that contribute. For certain values of $m$ one can have the perturbative contribution of equation (5.12) from the homogeneous solution. The particular solution must contain the terms that arise 
from  the right hand side of equation (4.4) but it may also contain terms that are the homogeneous solution to the equation in the perturbative limit, as we have just seen for the case $m=7$. To extend these consideration to automorphic forms associated with terms that have higher numbers of space-time derivatives one needs to know the   coefficient $A$. This is in general not known. However,  if we take $n=0$ in equation (4.12) we find that in ten dimensions 
$$
A= -{1 \over 16}(k-10)(9k-78)=  ({3p-1\over 2})({3p-3\over 2})= -s(s-1)
\eqno(5.17)$$
where $k=8+2p$, $s= {3p-1\over 2}$ or $s= {3-3p \over 2}$ for the term that can be written as $\partial^{2p} R^4$. Assuming this to be correct, we recognise this value as being of the form that allows a homogeneous solution to the equation in the perturbative limit and this leads in string frame to the contributions  
$$
e^{-p\phi}+ e^{2(p-1)\phi}
\eqno(5.18)$$
Clearly for $p>2$ the first term is not an allowed string correction, but the second term is always allowed and can be thought of as a $p$ loop contribution to the particular solution arising from the homogeneous solution to the Poisson equation in the perturbative limit as happened for the automorphic form associated with $\partial^6 R^4$. One can continue in the same vein to consider the perturbative contributions to terms with higher numbers of space-time derivatives. For example for the $R^9$ term we find, using equation (4.9),  that the right hand side of the Poisson equation it obeys can contain $\Phi_{E_{n+1}}^{R^4}\Phi_{E_{n+1}}^{\partial^4R^4}$.  Using the argument given above leads to the perturbative contributions 
$$
e^{-2\phi}(e^{0\phi}+e^{2\phi})(e^{0\phi}+e^{4\phi})= e^{-2\phi}(e^{0\phi}+
e^{2\phi}+e^{4\phi}+e^{6\phi})
\eqno(5.19)$$
leading to a tree, one loop, two loop and three loop contributions. This corresponds to the sum $9=3+5+1$. However, we can also have a homogeneous solution in the perturbative limit which, using equation (5.18) contributes 
$e^{8\phi}$ and so an additional four loop contribution. However, one can also write $9= 4+4+1$ this corresponds to a right hand side that contains the square of the automorphic form corresponding to the term $\partial^2 R^4$ which is mentioned below. As such the result just described could be modified by considering this term. Nonetheless this discussion illustrate that once the terms that can enter are better understood it may well be possible to derive which perturbative contribution occur in a simple way.

\bigskip
\noindent
{\bf {6 Discussion}}
\bigskip
In this paper we have studied the behaviour of string theory in all possible limits of its parameters. The higher derivative string corrections in $d=10-n$ dimensions are determined by automorphic forms of $E_{n+1}$. As such we have studied the behaviour of these automorphic forms and in particular the 
Poisson equations that they are thought to satisfy. Important for this derivation  was the identification of the string parameters in terms of the parameters that appear in the group  elements from which the automorphic forms are constructed.  

If one knew all the automorphic forms that occur then one would know all string and brane corrections in ten and lower dimensions which are related by toroidal dimensional reduction. However, for terms with more that 14 space-time derivatives very little is known. Unfortunately this is a highly technical subject in which it is difficult to make significant progress.

Using the results we have found for the behaviour of the Laplacian and in particular its  decompactification by one dimension we have  investigated  the Poisson equations that the automorphic forms are thought to obey. By making two simple assumptions we are able to derive the equations satisfied by the automorphic forms for all terms with less that 14 space-time derivatives and so derive, in a simple way,  much of what is known about these automorphic forms.  It would be interesting if one could derive rather than assume the form of the decompactification limit of the automorphic forms in equation (4.5).  We note that this expansion has a relatively simple form and this could indicate that there is a relatively straightforward derivation.  A true knowledge of the expansion would be most useful in applying the techniques of this paper to discover the properties of the automorphic forms in the effective action beyond 14 space-time derivatives. At first sight it would seem to be  straightforward to apply these  techniques  to terms with more than 14  space-time derivatives. However, there are two problems. 

The reader may have noticed that we did not consider a term in the effective action of the form $\partial^2 R^4$. This term does not appear in the effective action as it has a momentum prefactor that vanishes on shell. However, this term has been discussed in several places in the literature [12,25], but it is still not that well understood. It is thought that it should  appear in the supersymmetry arguments used to derive the terms that appear on the right hand side of the Poisson equation. This suggests that this term possess a corresponding automorphic form which  could appear among the product of automorphic forms on the right hand side of the Poisson equation. In principle one could write down its decompactification limit and then proceed in the way explained above.

There is however another possible complication for the terms with higher numbers of space-time derivatives. For these terms it is thought [12,25] that the automorphic form that appears in the effective action are themselves sums of automorphic forms that also obey individual Poisson equations.  One could  however, still hope to apply our techniques in that the automorphic forms that appear in the sum can each have a decompactification limit, as in  equation (4.5),  and this can be used in the Poisson equation that it satisfies. It would be interesting to take these two points into account and apply the method presented in this paper. 

In section five we discussed the perturbative limit in ten dimensions, but one could apply the same techniques to study the perturbative limit in less than ten dimensions and also  all the other limits in any dimension. It could be educational to carry this out. In particular it would be interesting to find the form of these expansions and in particular the powers of the parameters that can occur. For example in the limit of string perturbation theory one finds powers of $g_d^{-2+2q}$ where $q$ is the genus, but for example in the M theory limit what powers of $V_M$ can occur. 
It can be hoped that one might find further restriction on the automorphic forms in this way.

\medskip
\noindent
{\bf {Appendix A}}
\bigskip
In this section we will give an expression for the  Laplacian $\Delta$ on $E_{n+1}/H$ in terms of the parameters that we have used to parameterise the $E_{n+1}$ group element. In our application the parameters depend on the space-time and so are fields, those associated with the Cartan subalgebra were parameterised in this paper by the so called  Chevalley fields.  Using the relation between the parameters of the $d=10-n$-dimensional theory and the Chevalley fields, discussed earlier in this paper,  we will then be able to compute the Laplacian in the various limits in the subsequent appendices. 
\par
The $E_{n+1}/H$ Laplacian may be defined in terms of the components of the  metric on the $E_{n+1}/H$ symmetric space which we may write as 
$$
ds^{2} = {1 \over 2} \gamma_{ij} d \sigma^{i} d \sigma^{j}.
\eqno(A.1)
$$
where $\sigma_{i}$ are the parameters, or scalar fields,  parameterising the $E_{n+1}/H$ coset. The metric can be written in terms of the veilbein and for a coset space, such as $E_{n+1}/H$, the veilbein is contained in the Cartan forms  of  the group $E_{n+1}$.  The latter are given by  $g^{-1}dg$ where $g\in E_{n+1}$ but are subject to the  transformations $g\to gh$ with $h\in H$ which implements the equivalence relation concerning elements in the same coset. The group action on the coset is given by $g\to g_0g$ with $g_0\in E_{n+1}$. By writing the group element $g$  in its Iwasawa decomposition it is easy to see that we may use the $h$ transformations to bring the group element to the form 
$$
g=e^{\sum_{\vec{\alpha}}\chi_{\vec{\alpha}} E_{\vec{\alpha}}}e^{\vec{\varphi}.\vec{H}}
\quad {\rm with  \ the\  inverse\  given\  by } \quad 
g^{-1}=e^{-\vec{\varphi}.\vec{H}}e^{-\sum_{\vec{\alpha}}\chi_{\vec{\alpha}} E_{\vec{\alpha}}}
\eqno(A.2)
$$
where $\vec{H}$ and $E_{\vec{\alpha}}$ are the Cartan generators and positive root generators of $E_{n+1}$ and $\vec{\varphi}$, and $\chi_{\vec{\alpha}}$  are associated parameters, in fact  fields,  of the group element $g$. In terms of our notation above $\sigma^i=\{\vec{\varphi} , \chi_{\vec{\alpha}}\}$. 
\par
The algebra $E_{n+1}$, like all Kac-Moody algebras, possess an involution $I_c$ called the Cartan involution which acts on the generators as 
$$
I_c:(E_{\vec{\alpha}},E_{-\vec{\alpha}}, H)\to -(E_{-\vec{\alpha}},E_{\vec{\alpha}}, H),
\eqno(A.3)
$$
for ${\vec{\alpha}}$ a positive root. We may divide the generators of $E_{n+1}$ into those that are even,  that is $E_{\vec \alpha} - E_{-\vec\alpha}$,  and those that are odd, that is, $E_{\vec \alpha} + E_{-\vec \alpha}, \vec H$. The subgroup $H$ is generated by the even generators. Using this involution we may divide the Cartan form into its even and odd part by writing  
$$
g^{-1}dg= {\cal S}+ {\cal Q}
\eqno(A.4)
$$
where the odd  part of the Cartan form ${\cal S}$ is given by 
$$
{\cal S}={1 \over 2} \left( g^{-1}dg - I_c(g^{-1}dg)  \right)
\eqno(A.5)
$$
and the even   part ${\cal Q}$ is 
$$
{\cal Q}={1 \over 2} \left( g^{-1}dg + I_c(g^{-1}dg)  \right)
\eqno(A.6)
$$
The veilbein on the coset is the part of the Cartan form  in the coset direction,  that is the part that is odd,  and so in the quantity ${\cal S}$  given by equation (A.5). As a result the metric on the coset $E_{n+1}/H$  may be written as
$$
ds^{2}=Tr({\cal{S}} {\cal{S}})
\eqno(A.7)
$$
where we take the generators to be in some matrix representation. Using  the group element $g$ defined as in equation (A.2) one finds the Cartan form is given by
$$
g^{-1}dg =e^{-\vec{\varphi}.\vec{H}}e^{-\sum_{\vec{\alpha}} \chi_{\vec{\alpha}} E_{\vec{\alpha}}  } \left( \sum_{\vec{\alpha}} d \chi_{\vec{\alpha}} E_{\vec{\alpha}}  \right) e^{\sum_{\vec{\alpha}} \chi_{\vec{\alpha}} E_{\vec{\alpha}}  } e^{\vec{\varphi}.\vec{H}} + d \vec{\varphi}. \vec{H}
$$
$$
=e^{-\vec{\varphi}.\vec{H}} \left( \sum_{\vec{\alpha}} d \chi_{\vec{\alpha}} E_{\vec{\alpha}}  \right) e^{\vec{\varphi}.\vec{H}} + d \vec{\varphi}. \vec{H}+ O(\chi_{\vec \alpha}^2 )
$$
$$
=\sum_{\vec{\alpha}} d \chi_{\vec{\alpha}} \left( E_{\vec{\alpha}} + [-\vec{\varphi}.\vec{H},E_{\vec{\alpha}}] + {1 \over 2!} [ -\vec{\varphi}.\vec{H}, [-\vec{\varphi}.\vec{H},E_{\vec{\alpha}}] ] + ...  \right)  + d \vec{\varphi}.\vec{H}+ O(\chi_{\vec \alpha}^2 )
$$
$$
=\sum_{\vec{\alpha}} e^{-\vec{\varphi}\cdot\vec{\alpha} } d \chi_{\vec{\alpha}} E_{\vec{\alpha}} + d \vec{\varphi}.\vec{H}+ O(\chi_{\vec \alpha}^2 ),
\eqno(A.8)
$$
In what follows we will also be neglecting neglecting higher order terms in $\chi_{\vec{\alpha}}$ but we will not write this explicitly .  Writing $\vec{\alpha}=\sum_{a=1}^{n+1} n_{a} \alpha_{a}$ and $\underline \varphi = {2  \alpha _a \dot{\varphi}_a \over (\alpha _a , \alpha _a )} $ , where $\alpha_{a}$  are the simple roots of $E_{n+1}$ and $n_{a}$ are integer coefficients,  one has 
$$
g^{-1}dg = \sum_{\vec{\alpha}} e^{-\sum_{a,b} \dot{\varphi}_{a}A_{ab}n_{b} } d \chi_{\vec{\alpha}} E_{\vec{\alpha}} + d \vec{\varphi}.\vec{H},
\eqno(A.9)
$$
where $A_{ab}$ are the components of the Cartan matrix of $E_{n+1}$.   

The odd part of the Cartan form ${\cal S}$ under the Cartan involution $I_c$ is given by 
$$
{\cal S}
=\sum_{\vec{\alpha}} e^{-\sum_{a=1}^{n+1} \sum_{b=1}^{n+1} \dot{\varphi}_{a}A_{ab}n_{b} } d \chi_{\vec{\alpha}} \left( { E_{\vec{\alpha}} + E_{-\vec{\alpha}} \over 2} \right) + d \vec{\varphi}.\vec{H},
\eqno(A.10)
$$
The sum over $ \vec{\alpha}$ is over all positive roots which translates into a corresponding sum over $n_a$. 
\par
Using equations (A.7) and (A.10),  the components of the metric $\gamma_{ij}$ on the $E_{n+1}/H$ coset  space may then be found in terms of the Chevalley fields $\dot{\varphi}_{a}$, $a=1,...,n+1$ and the axions $\chi_{\vec{\alpha}}$ parameterising the group element $g \in E_{n+1}/H$ to be given by 
$$
ds^{2} = {1 \over 2} \gamma_{ij} d \sigma^{i} d \sigma^{j}=Tr({\cal{S}} {\cal{S}})= \sum_{a=1}^{n+1} \sum_{b=1}^{n+1} A_{ab} d \dot{\varphi}_{a} d \dot{\varphi}_{b} + {1 \over 2} \sum_{\vec{\alpha}}e^{-2\sum_{a=1}^{n+1}\sum_{b=1}^{n+1} \dot{\varphi}_{a} A_{ab} n_{b}} d \chi_{\vec{\alpha}} d \chi_{\vec{\alpha}}
\eqno(A.11)
$$
where we have taken 
$$
tr(H_{a} H_{b})=A_{ab}, \quad tr(E_{\vec{\alpha}} E_{-\vec{\beta}}) = tr(E_{-\vec{\alpha}} E_{\vec{\beta}}) = \delta_{ {\vec{\alpha}} {\vec{\beta}} }.
\eqno(A.12)
$$
One finds
$$
\gamma_{\dot{\varphi}_{a} \dot{\varphi}_{b} }= 2 A_{ab}, \quad \gamma_{\chi_{\vec{\alpha}} \chi_{\vec{\alpha}}}= e^{-2\sum_{a=1}^{n+1}\sum_{b=1}^{n+1} \dot{\varphi}_{a} A_{ab} n_{b}},
\eqno(A.13)
$$
and all other components of $\gamma_{ij}$ are zero.  
The components of the inverse metric $(\gamma^{-1})^{ij}$ are given by
$$
(\gamma^{-1})^{\dot{\varphi}_{a} \dot{\varphi}_{b} }= {1 \over 2} \left( A^{-1} \right)^{ab} \quad (\gamma^{-1})^{\chi_{\vec{\alpha}} \chi_{\vec{\alpha}}}= e^{2\sum_{a=1}^{n+1}\sum_{b=1}^{n+1} \dot{\varphi}_{a} A_{ab} n_{b}},
\eqno(A.14)
$$
and all other components $\gamma^{ij}$ are zero in the approximation we are taking.
\par
The Laplacian on the $E_{n+1}/H$ symmetric space is given by
$$
\Delta= {1 \over {\sqrt{ \gamma}}} \partial_{i} \left( \sqrt{\gamma} \gamma^{ij} \partial_{j} \right),
\eqno(A.15)
$$
where $\gamma_{ij}$ are the components of the metric on $E_{n+1}/H$,  $\gamma^{ij}$ are the components of the inverse metric and $\gamma=det(\gamma_{ij})$.  Substituting the components of the inverse metric in equations (A.10) into (A.11) and using $\gamma=2^{n+1}det(A_{ab})e^{-2\sum_{\vec{\alpha}}\sum_{a=1}^{n+1}\sum_{b=1}^{n+1} \dot{\varphi}_{a} A_{ab} n_{b}}$ one finds
$$
\Delta={1 \over 2} \sum_{a} \sum_{b} (A^{-1})^{ab} {\partial \over \partial \dot{\varphi}_{a}} {\partial \over \partial \dot{\varphi}_{b}}- {1 \over 2} \sum_{\vec{\alpha}} \sum_{b=1}^{n+1}  n_{b} {\partial \over \partial \dot{\varphi}_{b}} + \sum_{\vec{\alpha}}e^{2\sum_{a=1}^{n+1}\sum_{b=1}^{n+1} \dot{\varphi}_{a} A_{ab} n_{b}} {\partial^{2} \over \partial \chi_{\vec{\alpha}}}. 
\eqno(A.16)
$$
This expression may be simplified by noting that 
$$
{1 \over 2} \sum_{\vec{\alpha}} \sum_{b=1}^{n+1} n_{b} {\partial \over \partial \dot{\varphi}_{b}} = {1 \over 2} \sum_{\vec{\alpha}} \sum_{a=1}^{n+1} \sum_{b=1}^{n+1} \delta^{b}_{a} n_{a} {\partial \over \partial \dot{\varphi}_{b}}
={1 \over 2} \sum_{\vec{\alpha}} \sum_{a=1}^{n+1} \sum_{b=1}^{n+1} n_{a} \vec{\alpha}_{a}.\vec{\Lambda}_{b}  {\partial \over \partial \dot{\varphi}_{b}}
$$
$$
=\vec{\rho}.\sum_{b=1}^{n+1} \vec{\Lambda}_{b} {\partial \over \partial \dot{\varphi}_{b}}
=\sum_{a=1}^{n+1} \vec{\Lambda}_{a} . \sum_{b=1}^{n+1} \vec{\Lambda}_{b} {\partial \over \partial \dot{\varphi}_{b}}
=\sum_{a=1}^{n+1} \sum_{b=1}^{n+1} (A^{-1})^{ab}   {\partial \over \partial \dot{\varphi}_{b}}.
\eqno(A.17)
$$
The Laplacian may then be written as 
$$
\Delta={1 \over 2}\sum_{a} \sum_{b} (A^{-1})^{ab} {\partial \over \partial \dot{\varphi}_{a}} {\partial \over \partial \dot{\varphi}_{b}}- \sum_{a=1}^{n+1} \sum_{b=1}^{n+1} (A^{-1})^{ab}   {\partial \over \partial \dot{\varphi}_{b}} + \sum_{\vec{\alpha}}e^{2\sum_{a=1}^{n+1}\sum_{b=1}^{n+1} \dot{\varphi}_{a} A_{ab} n_{b}} {\partial^{2} \over \partial \chi_{\vec{\alpha}}^{2}}. 
\eqno(A.18)
$$
\medskip
\noindent
{\bf Appendix B: Limits of the $E_{n+1}$ Laplacian}
\medskip
As we have discussed taking  the parameters of the $d=10-n$ dimensional string theory to certain limits corresponds  to taking certain limits in the Chevalley fields $\dot{\varphi}_{i}$, $i=1,...,n+1$.  We will demonstrate this  process  by deriving the behaviour of the $E_{n+1}/H$ Laplacian in  the large volume limit of the M-theory torus $V_{M} \rightarrow \infty$ in the first subsection in detail and then give the behaviour of  the Laplacian in all limits in the other subsections. 
\medskip
\noindent
{\bf B.1 M-theory limit}
\medskip
The M-theory limit is the large volume limit of the M-theory torus $V_{n+1(M)}=e^{{8-n \over 3} \dot{\varphi}_{n+1}} \rightarrow \infty$.  As we will show, this limit results in the breaking of the $E_{n+1}$ algebra into a $GL(1) \times SL(n+1)$ subalgebra.  To analyse the M-theory limit we delete node $n+1$ in the Dynkin diagram given below.
$$
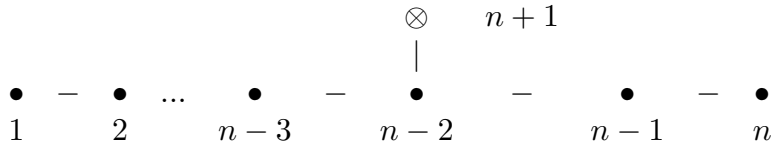
\matrix {
&&&&&  & \otimes & n+1&&\cr
&&&&&  & | &&& \cr
\bullet & - & \bullet & ... & \bullet & - & \bullet & - &\bullet & - & \bullet \cr
1&&2&&n-3 & & n-2 & & n-1 &  & n}
$$
\medskip
\centerline {The $E_{n+1}$ Dynkin diagram }
\medskip
Deleting node $n+1$ allows us to decompose the $E_{n+1}$ algebra in terms of the $GL(1) \times SL(n+1)$ subalgebra.  In this decomposition the simple roots of $E_{n+1}$ may be written 
$$
\vec{\alpha}_{i} = ( 0 , \underline{\alpha}_{i}), \quad i=1,2,...,n,
$$
$$
\vec{\alpha}_{n+1} = (x, -\underline{\lambda}_{n-2} )
\eqno(B.1)
$$
where $\underline{\alpha}_{i}$ and $\underline{\lambda}_{i}$ are the simple roots and fundamental weights of $SL(n+1)$ and $x^{2}= {8-n \over n+1}$.  The corresponding fundamental weights are
$$
\vec{\Lambda}^{i} = \left(  { \underline{\lambda}_{i}. \underline{\lambda}_{n-2} \over x^{2} }, \underline{\lambda}_{i}  \right), \quad i=1,2,...n,
\quad
\vec{\Lambda}^{n+1}=\left( {1 \over x}, \underline{0} \right).
\eqno(B.2)
$$
In deriving these and other such results in this paper we are using the techniques of reference [32], which the reader can consult for this method. 
\par
To preserve the $SL(n+1)$ part of the $GL(1) \times SL(n+1)$ subalgebra resulting from taking the $ V_{M} \rightarrow \infty $ limit, we must fix the quantities 
$$
\sum_{i=1}^{n}\dot{\varphi}_{i} \underline{\alpha}_{i} - \dot{\varphi}_{n+1} \underline{\lambda}_{n-2}.
\eqno(B.3)
$$
For further details on this point see section 4.1.4 of reference [29]. 
Defining the fields $\tilde{\varphi}_{j}$, $j=1,...,n$ by 
$$
\tilde{\varphi}_{j}= ( \sum_{i=1}^{n}\dot{\varphi}_{i} \underline{\alpha}_{i} - \dot{\varphi}_{n+1} \underline{\lambda}_{n-2}).\underline{\lambda}_{j}
=\dot{\varphi}_{j} - \underline{\lambda}_{j}.\underline{\lambda}_{n-2} \dot{\varphi}_{n+1}
=\dot{\varphi}_{j} - (\underline{A}^{-1})^{j \ n-2} \dot{\varphi}_{n+1},
\eqno(B.4)
$$
where $\underline{A}^{-1}$ is the inverse $SL(n+1)$ Cartan matrix, we see  that taking the $ V_{M(m)} =e^{{8-n \over 3} \dot{\varphi}_{n+1}} \rightarrow \infty $ limit is equivalent to taking the $\dot{\varphi}_{n+1}$ limit while holding $\tilde{\varphi}_{i}$, $i=1,2,...n$ fixed.
\par
To implement this limit we will  now rewrite the $E_{n+1}$ Laplacian in terms of the fields $\tilde{\varphi}_{i}$, $i=1,2,...,n$ and $\dot{\varphi}_{n+1}$ appropriate to the $V_{M} \rightarrow \infty$ limit.  The derivatives with respect to the Cartan subalgebra fields $\dot{\varphi}_{i}$, $i=1,2,...,n$ and $\dot{\varphi}_{n+1}$ become
$$
{ \partial \over \partial \dot{\varphi}_{i} }= {\partial \tilde{\varphi}_{k} \over \partial \dot{\varphi}_{i} } { \partial \over \partial \tilde{\varphi}_{k} } = { \partial \over \partial \tilde{\varphi}_{i} }, \quad i=1,2,...,n, 
\quad \quad 
{ \partial \over \partial \dot{\varphi}_{n+1} }= - \sum_{i=1}^{n} (\underline{A}^{-1})^{i \ n-2} { \partial \over \partial \tilde{\varphi}_{i} } + { \partial \over \partial \dot{\varphi}_{n+1} }
\eqno(B.5)
$$
and 
$$
{ \partial \over \partial \dot{\varphi}_{i} } { \partial \over \partial \dot{\varphi}_{j} } = { \partial \over \partial \tilde{\varphi}_{i} } { \partial \over \partial \tilde{\varphi}_{j} }, \quad i,j=1,2,...,n,
$$
$$
{ \partial \over \partial \dot{\varphi}_{i} } { \partial \over \partial \dot{\varphi}_{n+1} } = \left(-\sum_{j=1}^{n}\underline(A^{-1})^{j \ n-2} { \partial \over \partial \tilde{\varphi}_{j} } + { \partial \over \partial \dot{\varphi}_{n+1} } \right) { \partial \over \partial \tilde{\varphi}_{i} },
$$
$$
{ \partial \over \partial \dot{\varphi}_{n+1} } { \partial \over \partial \dot{\varphi}_{n+1} } = \left(-\sum_{i=1}^{n}\underline(A^{-1})^{i \ n-2} { \partial \over \partial \tilde{\varphi}_{i} } + { \partial \over \partial \dot{\varphi}_{n+1} } \right) \left(-\sum_{j=1}^{n}\underline(A^{-1})^{j \ n-2} { \partial \over \partial \tilde{\varphi}_{j} } + { \partial \over \partial \dot{\varphi}_{n+1} } \right).
\eqno(B.6)
$$
The inverse $E_{n+1}$ Cartan matrix $A^{-1}$, that appears in the Laplacian,  can be written in terms of  the inverse SL(n+1) Cartan matrix  $\underline{A}^{-1}$ as follows 
$$
(A^{-1})^{ij}= { (\underline{A}^{-1})^{i \ n-2} (\underline{A}^{-1})^{j \ n-2} \over x^{2}  } + (\underline{A}^{-1})^{i \ j}, \quad i,j=1,2,...,n, 
$$ 
$$
(A^{-1})^{i \ n+1 }= { (\underline{A}^{-1})^{i \ n-2} \over x^{2} }, \quad i=1,2,...,n,
$$
$$
(A^{-1})^{n+1 \ n+1 }= { 1 \over x^{2} },
\eqno(B.7)
$$
This result follows from  taking the inner product between the fundamental weights of $E_{n+1}$ decomposed with respect to node $n+1$, as given in equation (B.2).
Substituting these expressions for the derivatives with respect to the Cartan subalgebra and the decomposition of the inverse $E_{n+1}$ Cartan matrix with respect to node $n+1$ one finds 
$$
\Delta=  {1 \over 2} \sum_{i=1}^{n} \sum_{j=1}^{n} (A^{-1})^{ij} { \partial \over \partial \dot{\varphi}_{i} } { \partial \over \partial \dot{\varphi}_{j} } + \sum_{i=1}^{n} (A^{-1})^{i \ n+1 } { \partial \over \partial \dot{\varphi}_{i} } { \partial \over \partial \dot{\varphi}_{n+1} } + {1 \over 2} (A^{-1})^{n+1 \ n+1 } { \partial \over \partial \dot{\varphi}_{n+1} } { \partial \over \partial \dot{\varphi}_{n+1} }
$$
$$
 + \sum_{i=1}^{n} \sum_{j=1}^{n} (A^{-1})^{ij} { \partial \over \partial \dot{\varphi}_{j} } + \sum_{j=1}^{n} (A^{-1})^{ n+1 \ j} { \partial \over \partial \dot{\varphi}_{j} } + \sum_{i=1}^{n} (A^{-1})^{i \ n+1 } { \partial \over \partial \dot{\varphi}_{n+1} } + (A^{-1})^{n+1 \ n+1 } { \partial \over \partial \dot{\varphi}_{n+1} }
$$
$$
  + \sum_{\vec{\alpha}}e^{2\sum_{a=1}^{n+1}\sum_{b=1}^{n+1} \dot{\varphi}_{a} A_{ab} n_{b}} {\partial^{2} \over \partial \chi_{\vec{\alpha}}^{2}}
$$  
$$
= {1 \over 2} \sum_{i=1}^{n} \sum_{j=1}^{n} \left(   { (\underline{A}^{-1})^{i \ n-2} (\underline{A}^{-1})^{j \ n-2} \over x^{2}  } + (\underline{A}^{-1})^{i \ j}  \right) { \partial \over \partial \tilde{\varphi}_{i} } { \partial \over \partial \tilde{\varphi}_{j} } 
$$
$$
+  \sum_{i=1}^{n} { (\underline{A}^{-1})^{i \ n-2} \over x^{2} } \left(-\sum_{j=1}^{n}\underline(A^{-1})^{j \ n-2} { \partial \over \partial \tilde{\varphi}_{j} } + { \partial \over \partial \dot{\varphi}_{n+1} } \right) { \partial \over \partial \tilde{\varphi}_{i} }
$$
$$
+ {1 \over 2x^{2} } \left(-\sum_{i=1}^{n}\underline(A^{-1})^{i \ n-2} { \partial \over \partial \tilde{\varphi}_{i} } + { \partial \over \partial \dot{\varphi}_{n+1} } \right) \left(-\sum_{j=1}^{n}\underline(A^{-1})^{j \ n-2} { \partial \over \partial \tilde{\varphi}_{j} } + { \partial \over \partial \dot{\varphi}_{n+1} } \right)
$$
$$
+ \sum_{i=1}^{n} \sum_{j=1}^{n} \left(   { (\underline{A}^{-1})^{i \ n-2} (\underline{A}^{-1})^{j \ n-2} \over x^{2}  } - (\underline{A}^{-1})^{i \ j}  \right) { \partial \over \partial \tilde{\varphi}_{j} } - \sum_{j=1}^{n} { (\underline{A}^{-1})^{j \ n-2} \over x^{2} } { \partial \over \partial \tilde{\varphi}_{j} } 
$$
$$
- \sum_{i=1}^{n} { (\underline{A}^{-1})^{i \ n-2} \over x^{2} } \left( - \sum_{i=1}^{n} (\underline{A}^{-1})^{i \ n-2} { \partial \over \partial \tilde{\varphi}_{i} } + { \partial \over \partial \dot{\varphi}_{n+1} } \right) - {1 \over x^{2} } \left(- \sum_{i=1}^{n} (\underline{A}^{-1})^{i \ n-2} { \partial \over \partial \tilde{\varphi}_{i} } + { \partial \over \partial \dot{\varphi}_{n+1} } \right) 
$$
$$
+ \sum_{\vec{\alpha}}e^{2\sum_{a=1}^{n+1}\sum_{b=1}^{n+1} \dot{\varphi}_{a} A_{ab} n_{b}} {\partial^{2} \over \partial \chi_{\vec{\alpha}}^{2}}
$$
$$
= {1 \over 2x^{2} } { \partial \over \partial \dot{\varphi}_{n+1} } { \partial \over \partial \dot{\varphi}_{n+1} }  - {(3n^{2} - n -4) \over 2(8-n) } { \partial \over \partial \dot{\varphi}_{n+1} } + {1 \over 2} \sum_{i=1}^{n} \sum_{j=1}^{n} (\underline{A}^{-1})^{i \ j}{ \partial \over \partial \tilde{\varphi}_{i} } { \partial \over \partial \tilde{\varphi}_{j} }  -  \sum_{i=1}^{n} \sum_{j=1}^{n} (\underline{A}^{-1})^{i \ j} { \partial \over \partial \tilde{\varphi}_{j} } 
$$
$$
  + \sum_{\vec{\alpha}}e^{2\sum_{a=1}^{n+1}\sum_{b=1}^{n+1} \dot{\varphi}_{a} A_{ab} n_{b}} {\partial^{2} \over \partial \chi_{\vec{\alpha}}^{2}}.
\eqno(B.8)
$$
In deriving this last equation  we have used that 
$ \underline{\lambda}_{i}. \underline{\lambda}_{j}={i(n+1 -j) \over n+1}$, for $i \leq j$ and $\sum_{i=1}^{k} = { k(k+1) \over 2}$. 
\par
The derivatives with respect to the axionic terms in the Laplacian, given equation (B.8), possess the coefficient $e^{2\sum_{a=1}^{n+1} \sum_{b=1}^{n+1} \dot{\varphi}_{a} A_{ab} n_{b}}$. Writing the fields $\dot{\varphi}_{a}$ in the basis $\tilde{\varphi}_{a}$ given in equation (B.4) appropriate for taking the $V_{M(m)} \rightarrow \infty$ limit we find
$$
e^{2\sum_{a=1}^{n+1} \sum_{b=1}^{n+1} \dot{\varphi}_{a} A_{ab} n_{b}}=e^{2\sum_{a=1}^{n} \sum_{b=1}^{n} \tilde{\varphi}_{a} \underline{A}_{ab} n_{b} - 2 \tilde{\varphi}_{n-2}n_{n+1}  + 4 \dot{\varphi}_{n+1} n_{n+1}}.
$$
$$
=e^{2\sum_{a=1}^{n} \sum_{b=1}^{n} \tilde{\varphi}_{a} \underline{A}_{ab} n_{b}  - 2 \tilde{\varphi}_{n-2}n_{n+1}} V_{M(m)}^{\left( 12 \over 8-n \right) n_{n+1}}
\eqno(B.9)
$$
In the $V_{M(m)} \rightarrow \infty$ limit, the derivatives with respect to the axions associated with positive roots containing the simple root $\vec{\alpha}_{n+1}$, and therefore having $n_{n+1} >0$, appear to diverge.  However, this is a consequence of the Laplacian being constructed from components of the inverse group metric $\gamma^{ij}$.  To examine the behaviour of the group metric in the $V_{M(m)} = e^{{8-n \over 3} \dot{\varphi}_{n+1}} \rightarrow \infty$ limit we rewrite the coefficients of the axions in terms of the fields relevant to the  $V_{M(m)} \rightarrow \infty$ limit
$$
ds^{2}_{E_{n+1}} = {1 \over 2} \gamma_{ij} d \sigma^{i} d \sigma^{j}=Tr({\cal{S}} {\cal{S}})= \sum_{a=1}^{n+1} \sum_{b=1}^{n+1} A_{ab} d \dot{\varphi}_{a} d \dot{\varphi}_{b} + {1 \over 2} \sum_{\vec{\alpha }>0}e^{-2\sum_{a=1}^{n+1}\sum_{b=1}^{n+1} \dot{\varphi}_{a} A_{ab} n_{b}} d \chi_{\vec{\alpha}} d \chi_{\vec{\alpha}}
$$
$$
= \sum_{a=1}^{n+1} \sum_{b=1}^{n+1} A_{ab} d \dot{\varphi}_{a} d \dot{\varphi}_{b} + {1 \over 2} \sum_{\vec{\alpha}>0}e^{-2\sum_{a=1}^{n} \sum_{b=1}^{n} \tilde{\varphi}_{a} \underline{A}_{ab} n_{b}  + 2 \tilde{\varphi}_{n-2}n_{n+1}} V_{M(m)}^{-\left( 12 \over 8-n \right) n_{n+1}}d \chi_{\vec{\alpha}} d \chi_{\vec{\alpha}}.
\eqno(B.10)
$$
In taking the $V_{M(m) \rightarrow \infty}$ limit we see that the axionic terms in the group metric $ds^{2}_{E_{n+1}}$ associated with positive roots containing the simple root $\vec{\alpha_{n+1}}$ vanish and therefore we are left with non-zero axionic metric components of the group $SL(n+1)$ rather than the full $E_{n+1}$ group. Put another way the sum in equation (B.9) no longer runs over roots that contain $\alpha_{n+1}$. Therefore in order to deduce the behaviour of the $E_{n+1}$ Laplacian in the $V_{M(m) \rightarrow \infty}$ limit we should first take the limit in the group metric, which leaves us with the group metric for the remaining $GL(1) \times SL(n+1)$ subgroup and then calculate the components of the group metric from which we calculate the Laplacian rather than take the limit directly in the Laplacian as was done above.  The components of the group metric associated with the Chevalley fields $\gamma_{\dot{\varphi}_{a} \dot{\varphi}_{b}}$ are unchanged in this limit and given in equation (A.9). The axionic components of the group metric in this limit are
$$
\gamma_{\chi_{\underline{\alpha}} \chi_{\underline{\alpha}} }= e^{-2\sum_{a=1}^{n} \sum_{b=1}^{n} \tilde{\varphi}_{a} \underline{A}_{ab} n_{b}}
\eqno(B.11)
$$
with the remaining components, including the axionic terms in the group metric $ds^{2}_{E_{n+1}}$ associated with positive roots containing the simple root $\vec{\alpha_{n+1}}$, being zero. We note that in this latter equation the objects $\tilde{\varphi}_{a} $ which are fixed in the limit appear. The axionic components of the inverse group metric in this limit are
$$
\gamma^{\chi_{\underline{\alpha}} \chi_{\underline{\alpha}} }= e^{2\sum_{a=1}^{n} \sum_{b=1}^{n} \tilde{\varphi}_{a} \underline{A}_{ab} n_{b}}
\eqno(B.12)
$$
with the remaining components, including the axionic terms in the group metric $ds^{2}_{E_{n+1}}$ associated with positive roots containing the simple root $\vec{\alpha_{n+1}}$, being zero.  Using equations (A.14) and (B.12) we find that  in the $V_{M(m)} \rightarrow \infty$ limit, the Laplacian is given by
$$
\Delta= {1 \over 2x^{2} } { \partial \over \partial \dot{\varphi}_{n+1} } { \partial \over \partial \dot{\varphi}_{n+1} } - {(3n^{2} - n -4) \over 2(8-n) } { \partial \over \partial \dot{\varphi}_{n+1} } + {1 \over 2} \sum_{i=1}^{n} \sum_{j=1}^{n} (A^{-1})^{ij} { \partial \over \partial \tilde{\varphi}_{i} } { \partial \over \partial \tilde{\varphi}_{j} }  
$$
$$
- \sum_{i=1}^{n} \sum_{j=1}^{n} (A^{-1})^{ab}   {\partial \over \partial \tilde{\varphi}_{b}} +\sum_{\vec{\alpha}}e^{2\sum_{a=1}^{n}\sum_{b=1}^{n} \tilde{\varphi}_{a} \underline{A}_{ab} n_{b}} {\partial^{2} \over \partial \chi^{2}_{\underline{\alpha}}}
$$
$$
= {1 \over 2x^{2} } { \partial \over \partial \dot{\varphi}_{n+1} } { \partial \over \partial \dot{\varphi}_{n+1} } - {(3n^{2} - n -4) \over 2(8-n) } { \partial \over \partial \dot{\varphi}_{n+1} } + \Delta_{SL(n+1)}.
\eqno(B.13)
$$

\medskip
\noindent
{ \bf {B.2 Type IIB Volume Limit}}
\medskip

In the type IIB limit $V_{n(B)} \rightarrow \infty$. Examining equation (2.5) we find that this corresponds to  deleting  node $n-1$ in the Dynkin diagram given below.
$$
\matrix {
&&&&&  & \bullet & n+1&&\cr
&&&&&  & | &&& \cr
\bullet & - & \bullet & ... & \bullet & - & \bullet & - &\otimes & - & \bullet \cr
1&&2&&n-3 & & n-2 & & n-1 &  & n}
$$
\medskip
\centerline {The $E_{n+1}$ Dynkin diagram }
\medskip
Deleting node n-1  decomposes the $E_{n+1}$ algebra in terms of  a $GL(1) \times SL(n) \times SL(2)$ subalgebra.  In this decomposition the simple roots of $E_{n+1}$ may be written, using the techniques of reference [31] as 
$$
\vec{\alpha}_{i}=\left( 0 , 0, \underline{\alpha}_{i} \right), \ \ \ i=1,...,n-2,
\quad 
\vec{\alpha}_{n-1}=\left( x , -\mu_{1}, - \underline{\lambda}_{n-2} \right), 
$$
$$
\vec{\alpha}_{n}=\left( 0, \beta_{1},0   \right), 
\quad
\vec{\alpha}_{n+1}=\left( 0 , 0, \underline{\alpha}_{n-1} \right), 
\eqno(B.14)
$$
where the underline denotes $SL(n)$ simple roots and fundamental weights and $\mu_{1}$, $\beta_{1}$ are the fundamental weight and simple root of $SL(2)$ respectively.  The variable $x$ is fixed by the condition on the length of the simple roots, $\vec{\alpha}_{n}^{2}=2=x^{2}+\underline{\lambda}_{n-2}^{2}+ \mu_{1}^{2}$, this leads to $x^{2}={8-n \over 2n}$.  The corresponding fundamental weights are
$$
\vec{\Lambda}^{i}=\left({\underline{\lambda}_{i}.\underline{\lambda}_{n-2} \over x}, 0, \underline{\lambda}_{i} \right), \ \ \ i=1,...,n-2, 
\quad 
\vec{\Lambda}^{n-1}=\left({1 \over x}, \underline{0} \right), 
$$
$$
\vec{\Lambda}^{n}=\left( {1 \over 2x}, \mu_{1}, \underline{0} \right),
\quad 
\vec{\Lambda}^{n+1}=\left({\underline{\lambda}_{n-1}.\underline{\lambda}_{n-2} \over x}, 0, \underline{\lambda}_{n-1} \right). 
\eqno(B.15)
$$
In taking the $ V_{n(B)} \rightarrow \infty $ limit, which is equivalent to $\varphi_{n-1} \rightarrow \infty$, we must fix the quantities 
$$
\tilde{\varphi}_{i}=\dot{\varphi}_{i} - (\underline{A}^{-1})^{i \ n-2} \dot{\varphi}_{n-1}, \quad i=1,...,n-2, 
$$
$$
\tilde{\varphi}_{n+1}=\dot{\varphi}_{n+1} - (\underline{A}^{-1})^{n-1 \ n-2} \dot{\varphi}_{n-1},
\quad 
\tilde{\varphi}={1 \over 2}\dot{\varphi}_{n-1}- \dot{\varphi}_{n}
\eqno(B.16)
$$
where $\underline{A}^{-1}$ is the inverse $SL(n)$ Cartan matrix, to preserve the $SL(2) \times SL(n)$ symmetry. 
\par
In the large volume limit of the type IIB torus $V_{n(B)}= e^{{8-n \over 4} \dot{\varphi}_{n-1} } \rightarrow \infty$ the Laplacian $\Delta$ becomes
$$
\Delta= { 1 \over 2x^{2} } { \partial \over \partial \dot{\varphi}_{n-1} } { \partial \over \partial \dot{\varphi}_{n-1} } - { (2n^{2} - n ) \over (8-n)   } { \partial \over \partial \dot{\varphi}_{n-1} } + \Delta_{SL(n)} + \Delta_{SL(2)}
$$
$$
= { n \over (8-n) } { \partial \over \partial \dot{\varphi}_{n-1} } { \partial \over \partial \dot{\varphi}_{n-1} } - { (2n^{2} - n ) \over (8-n)   }  { \partial \over \partial \dot{\varphi}_{n-1} } + \Delta_{SL(n)}+ \Delta_{SL(2)}
\eqno(B.17)
$$
where we have used $x^{2}= {8-n \over 2n}$.
\medskip
\noindent
{ \bf {B.3 Type IIA Volume Limit}}
\medskip
In the type IIA limit $V_{n(A)}= =e^{{8-n \over 8} (\dot{\varphi}_{n} + 2 \dot{\varphi}_{n+1})  }\rightarrow \infty$ and as a result it corresponds to deleting  nodes $n+1$ and $n$ in the Dynkin diagram given below.
$$
\matrix {
&&&&&  & \otimes & n+1&&\cr
&&&&&  & | &&& \cr
\bullet & - & \bullet & ... & \bullet & - & \bullet & - &\bullet & - & \otimes \cr
1&&2&&n-3 & & n-2 & & n-1 &  & n}
$$
\medskip
\centerline {The $E_{n+1}$ Dynkin diagram }
\medskip
Deleting nodes $n$ and $n+1$ of the Dynkin diagram leads to the  decomposition of  $E_{n+1}$ into the subalgebra  $ GL(1) \times GL(1) \times SL(n)$. As a result we will now examine how the roots and weights of $E_{n+1}$ decompose in terms of those of
$GL(1)\times GL(1) \times SL(n)$.
\par
Let us carry out the decomposition by  first
deleting  node
$n$ to find the roots and fundamental weights of $D_{n}$ and then
delete node
$n+1$ to find the algebra $SL(n)$. The simple roots of
$E_{n+1}$ can be expressed as
$$
\vec \alpha_{i} = \left(0, \tilde{\alpha}_{i} \right),
\quad i
= 1,...,n-1, n+1 \quad
\vec \alpha_{n}=\left(x,-  \tilde{\lambda}_{n-1} \right).
\eqno(B.18)
$$
Here
$\tilde{\alpha}_{i}, i =1,...,n$ are the roots of $D_
{n}$
and
$\tilde{\lambda}_i$ are its fundamental weights which are given
by
$$
\vec \Lambda_{i} = \left({\tilde{\lambda}_i\cdot \tilde{\lambda}_{n-1} \over x},
\tilde{\lambda}_i \right),
\quad
i = 1,...,n-1, n+1 \quad
\vec \Lambda_{n}=\left({1\over x}, \tilde{0} \right).
\quad\
\eqno(B.19)
$$
The variable
$x$ is fixed by demanding that $\vec{\alpha}_{n}^2=2=x^2+
\tilde{\lambda}_{n-1}^2$.
\par
We now delete node $n$ to find the $A_{n-1}$ algebra. The roots of
$E_{n+1}$ are found from the above roots by substituting the
corresponding
decomposition of the $D_{n}$ roots and weights into those of $A_
{n-1}$.
The roots of $D_{n}$ in terms of those of $A_{n-1}$ are given by
$\tilde{\alpha}_i = \left(0, \underline{\alpha}_{i}\right),\
i=1,...,n -1$ and
$\tilde{\alpha}_{n} = \left(y,- \underline
\lambda_{n-2}\right)$ while the fundamental weights are given by
$\tilde{\lambda}_i = \left({\lambda _{n-2}\cdot
\lambda_{i}\over y},
\underline{\lambda}_{i}\right)\ i=1,...,n -1$ and
$\tilde{\lambda}_{n+1} = \left({1\over y}, \underline{0} \right)$. Requiring
${\tilde \alpha}_{n+1}^2=2$ gives $y^2={4\over n}$
We then find that the roots of $E_{n+1}$ are given by
$$
\vec \alpha_{i} = \left(0, 0,\underline \alpha_{i} \right), \quad
i =1,...,n-1, 
\vec \alpha_{n} = \left(x , -{\lambda _{n-2}\cdot \lambda_{n-1}\over
y}, -\underline    \lambda_{n-1}
\right), 
\vec \alpha_{n+1} =\left(0,y, -\underline \lambda_{n-2} \right).
\eqno(B.20)
$$
The fundamental weights of $E_{n+1}$ are found in the same way to be
$$
\vec{\Lambda}_{i} = \left( {c_i\over x},{\lambda _{n-2}\cdot
\lambda_{i}\over y},
\underline \lambda_{i} \right),  \quad i =1,...,n-1, 
$$
$$
\vec{\Lambda}_{n} = \left( {1\over x},0 , \underline 0  \right),
\quad 
\vec{\Lambda}_{n+1} = \left( {n-2\over 4 x},{1\over y} , \underline 0
\right),
\eqno(B.21)
$$
where $ c_i={i\over 2}, \ i=1,\ldots ,n-2$ and  $ c_{n-1}={n \over 4}$.
As $\tilde{\lambda}_{n-1}^2={n\over 4}$ we find that $x^2={8-n\over 4}$.

\par
In taking the $ V_{n(A)} \rightarrow \infty $ limit, which is equivalent to $\varphi_{n-1} \rightarrow \infty$, we must fix the quantities 
$$
\tilde{\varphi}_{j}={\dot \varphi}_{j}  - (\underline{A}^{-1})^{i \ n-1} \dot{\varphi}_{n}  - (\underline{A}^{-1})^{i \ n-2} \dot{\varphi}_{n+1}
\eqno(B.22)
$$
where $\underline{\alpha}_{i}$ and $\underline{\lambda}_{i}$, $i=1,...,n-1$ are the simple roots and fundamental weights of $SL(n)$ respectively to preserve the $SL(n)$ symmetry and in addition fix 
$$
\dot{\varphi}_{g}=-{3 \over 2} \dot{\varphi}_{n} + \dot{\varphi}_{n+1} \quad 
\eqno(B.23)
$$
to preserve the type IIA string coupling. 
\par
In the large volume limit of the type IIA torus $V_{n(A)}=e^{{8-n \over 8} (\dot{\varphi}_{n} + 2 \dot{\varphi}_{n+1})  }$ the Laplacian $\Delta$ becomes
$$
\Delta= { 4n \over (8-n) } { \partial \over \partial \dot{\varphi}_{V} } { \partial \over \partial \dot{\varphi}_{V} } - \left( { 4n^{2} -2n  \over 8-n   } \right) { \partial \over \partial \dot{\varphi}_{V} } + { \partial \over \partial \dot{\varphi}_{g} } { \partial \over \partial \dot{\varphi}_{g} }  + { \partial \over \partial \dot{\varphi}_{g} } + \Delta_{SL(n)}
\eqno(B.24)
$$
where we have defined $\dot{\varphi}_{V}=\dot{\varphi}_{n} +2\dot{\varphi}_{n+1} $ and used $x^{2}= {8-n \over 2n}$.
\medskip
\noindent
{ \bf {B.4 Decompactification of a single dimension Limit}}
\medskip

Using equation (2.7) we see that  the decompactification of a single dimension limit ${r_{d+1} \over l_{d+1}} =e^{{8-n \over 9-n} \dot{\varphi}_{1}}\rightarrow \infty$ corresponds to the  deletion of  node $1$ in the Dynkin diagram given below.
$$
\matrix {
&&&&&  & \bullet & n+1&&\cr
&&&&&  & | &&& \cr
\otimes & - & \bullet & ... & \bullet & - & \bullet & - &\bullet & - & \bullet \cr
1&&2&&n-3 & & n-2 & & n-1 &  & n}
$$
\medskip
\centerline {The $E_{n+1}$ Dynkin diagram }
\medskip
Deleting node 1 decomposes the $E_{n+1}$ algebra in terms of the $GL(1) \times E_{n}$ subalgebra.  In this decomposition the simple roots of $E_{n+1}$ may be written
$$
\vec{\alpha}_{1}=\left( x , -\hat{\lambda}_{1} \right),  
\quad 
\vec{\alpha}_{i}=\left(0, \hat{\alpha}_{i-1} \right), \ \ \ i=2,...,n+1, 
\eqno(B.25) 
$$
where the hat denotes $E_{n}$ simple roots and fundamental weights.  The variable $x$ is fixed by the condition on the length of the simple roots, $\vec{\alpha}_{1}^{2}=2=x^{2}+\hat{\lambda}_{1}^{2}$.  The corresponding fundamental weights are
$$
\vec{\Lambda}^{1}=\left({1 \over x}, \underline{0} \right), \quad 
\vec{\Lambda}^{i}=\left({\hat{\lambda}_{i-1}.\hat{\lambda}_{1} \over x}, \hat{\lambda}_{i-1} \right), \ \ \ i=2,...,n+1. 
\eqno(B.26)
$$
We now proceed to calculate the inner products of the $E_{n}$ fundamental weights in order to calculate $x$ in terms of $n$.  To do this we decompose the $E_{n}$ algebra into a $GL(1) \times SL(n)$ subalgebra by deleting node $n+1$, one finds
$$
\hat{\alpha}_{i}=\left( 0 , \underline{\alpha}_{i} \right), \ \ \ i=1,...,n-1, 
\quad 
\hat{\alpha}_{n}= \left(y, -\underline{\lambda}_{n-3} \right), 
\eqno(B.27)
$$
with fundamental weights
$$
\hat{\lambda}_{i}=\left( {\underline{\lambda}_{i}. \underline{\lambda}_{n-3} \over y}, \underline{\lambda}_{i} \right), \ \ \ i=1,...,n-1, 
\quad 
\hat{\lambda}_{n}=\left( {1 \over y} , \underline{0}  \right). 
\eqno(B.28)
$$
The variable $y$ is fixed by the condition $\hat{\alpha}_{n-2}^{2}=2$, this gives $y^{2}={9-n \over n}$.  We then have
$$
\hat{\lambda}_{1}.\hat{\lambda}_{1}=\left({ 3 \over ny}, \underline{\lambda}_{1} \right).\left({3 \over ny}, \underline{\lambda}_{1} \right)  
={9 \over n^{2} y^{2} }+{n-1 \over n} 
={10-n \over 9-n},
\eqno(B.29)
$$
where we have made use of the expression $\underline{\lambda}_{i}.\underline{\lambda}_{j}={i\left( n-j \right) \over n}$ for $i \leq j$.
We may now substitute this back into $\vec{\alpha}_{1}.\vec{\alpha}_{1}$ to fix the variable $x$,
$$
x^{2}=2-\hat{\lambda}_{1}.\hat{\lambda}_{1} 
={8-n \over 9-n}.
\eqno(B.30)
$$

In taking the $ {r_{d+1} \over l_{d+1} } \rightarrow \infty $ limit, which is equivalent to $\varphi_{1} \rightarrow \infty$, we must fix the quantities 
$$
{\tilde{\varphi}}_{i-1}=
\dot{\varphi}_{i-1} - (\underline{A}^{-1})^{i-1 \ 1} \dot{\varphi}_{1} - \dot{\varphi}_{1} \underline{\lambda}_{1}
\eqno(B.31)
$$
for $i=2,...,n+1$ and where $(\underline{A}^{-1})$ is the inverse $E_{n}$ Cartan matrix, to preserve the $E_{n}$ symmetry.
\par

\par
In the decompactification of a single dimension limit ${r_{d+1} \over l_{d+1} }= e^{{8-n \over 9-n} \dot{\varphi}_{1} } \rightarrow \infty$ the Laplacian $\Delta$ becomes
$$
\Delta= { 1 \over 2x^{2} } { \partial \over \partial \dot{\varphi}_{1} } { \partial \over \partial \dot{\varphi}_{1} } - { (-n^{2} + 17n - 12) \over (2x^{2}(9-n))   } { \partial \over \partial \dot{\varphi}_{1} } + \Delta_{E_{n}}
$$
$$
= { (9-n) \over 2(8-n) } { \partial \over \partial \dot{\varphi}_{1} } { \partial \over \partial \dot{\varphi}_{1} } - { (-n^{2} + 17n - 12) \over 2(8-n)   } { \partial \over \partial \dot{\varphi}_{1} } + \Delta_{E_{n}}
\eqno(B.32)
$$
where we have used $x^{2}= {8-n \over 9-n}$.

\medskip
\noindent
{ \bf {B.5 $j$ dimensional subtorus limit}}
\medskip

The $j$ dimensional subtorus limit $V_{j} = ^{{8-n \over 8-n+j} \dot{\varphi}_{j} }
\rightarrow \infty$ corresponds to  deleting  node $j$ in the Dynkin diagram given below.
$$
\matrix {
&&&&&&  & \otimes & n+1&&\cr
&&&&&&  & | &&& \cr
\bullet & - & \bullet & ... & \otimes & ... & - & \bullet & - &\bullet & - & \bullet \cr
1&&2&&j && & n-2 & & n-1 &  & n}
$$
\medskip
\centerline {The $E_{n+1}$ Dynkin diagram }
\medskip
Deleting node $j$  decomposes the $E_{n+1}$ algebra in terms of a $GL(1) \times SL(j) \times E_{n+1-j}$ subalgebra.  In this decomposition the simple roots of $E_{n+1}$ may be written as 
$$
\vec{\alpha}_{i}=\left(0, \underline{\alpha}_{i}, \hat{0} \right), \ \ \ i=1,...,j-1,
\quad 
\vec{\alpha}_{j}=\left( x , -\underline{\lambda}_{j-1}, - \hat{\lambda}_{1} \right),  
$$
$$
\vec{\alpha}_{k}=\left(0, \hat{\alpha}_{k-j} \right), \ \ \ k=j+1,...,n+1, 
\eqno(B.33) 
$$
where the underline and the hat denote $SL(j)$ and $E_{n+1-j}$ quantities and $\alpha$, $\lambda$ are the respective simple root and fundamental weights of the corresponding algebra.  The corresponding fundamental weights are
$$
\vec{\Lambda}^{i}=\left(  {\underline{\lambda}_{i}.\underline{\lambda}_{j-1} \over x}, \underline{\lambda}_{i}, \hat{0} \right), \ \ \ i=1,...,j-1,
\quad 
\vec{\Lambda}^{j}=\left(  {1 \over x}, \underline{0}, \hat{0} \right),
$$
$$
\vec{\Lambda}^{k}=\left({\hat{\lambda}_{k-j}.\hat{\lambda}_{1} \over x}, \underline{0}, \hat{\lambda}_{k-j} \right), \ \ \ k=j+1,...,n+1. 
\eqno(B.34)
$$
The variable $x$ is fixed by the condition on the length of the simple roots, $\vec{\alpha}_{j}^{2}=2=x^{2}+\hat{\lambda}_{1}.\hat{\lambda}_{1}+\underline{\lambda}_{j-1}.+\underline{\lambda}_{j-1}$.  After some work one finds
$$
x^{2}={(n+1)(8-n+j) - 9j \over j(n+1-j)(8-n+j)}.
\eqno(B.35)
$$

In taking the $ V_{j} \rightarrow \infty $ limit, which is equivalent to $\varphi_{j} \rightarrow \infty$, we must fix the quantities 
$$
\tilde{{\varphi}}_{i} =\dot{\varphi}_{i} - \dot{\varphi}_{j} (\underline{A}^{-1})^{i \ j-1} 
\eqno(B.36)
$$
for $i=1,...,j-1$ and
$$
{\hat{\varphi}}_{n+1-i} ={\dot \varphi}_{i}  - \dot{\varphi}_{j} (\hat{A}^{-1})^{i \ 1}  \hat{\lambda}_{1}
\eqno(B.37)
$$ 
for $i=n+1-j,...,n+1$ to preserve the $SL(j) \times E_{n+1-j}$ symmetry where $(\underline{A}^{-1})$ is the inverse Cartan matrix of $SL(j)$ and $(\hat{A}^{-1})^{i \ 1}$ is the inverse Cartan matrix of $E_{n+1-j}$.
 
\par
In the large volume limit of the $j$ dimensional subtorus $V_{j}= e^{{8-n \over 8-n+j} \dot{\varphi}_{j} } \rightarrow \infty$ the Laplacian $\Delta$ becomes
$$
\Delta= { 1 \over 2x^{2}  } { \partial \over \partial \dot{\varphi}_{j} } { \partial \over \partial \dot{\varphi}_{j} } - { (-n^{2} + 16n - 8j +nj - 4) \over 2x^{2}(8-n+j)   } { \partial \over \partial \dot{\varphi}_{j} } + \Delta_{E_{n+1-j}}
$$
$$
= { j(n+1-j)(8-n+j) \over 2((n+1)(8-n+j) -9j) } { \partial \over \partial \dot{\varphi}_{j} } { \partial \over \partial \dot{\varphi}_{j} } 
$$
$$
- { j(n+1-j) \over 2((n+1)(8-n+j) -9j) } (-n^{2}+16n-8j+nj-4) { \partial \over \partial \dot{\varphi}_{j} } + \Delta_{E_{n+1-j}}
\eqno(B.38)
$$
where we have used $x^{2}= {(n+1)(8-n+j) - 9j \over j(n+1-j)(8-n+j)}$.

\medskip
\noindent
{ \bf {B.6 Perturbative Limit}}
\medskip

The $d$ dimensional perturbative limit $g_{d}=  =e^{-\left( {8-n \over 4} \right)}
\rightarrow 0$ corresponds to  deleting  node $n$ in the Dynkin diagram given below.
$$
\matrix {
&&&&&  & \bullet & n+1&&\cr
&&&&&  & | &&& \cr
\bullet & - & \bullet & ... & \bullet & - & \bullet & - &\bullet & - & \otimes \cr
1&&2&&n-3 & & n-2 & & n-1 &  & n}
$$\
\medskip
\centerline {The $E_{n+1}$ Dynkin diagram }
\medskip
Deleting node n  decomposes the $E_{n+1}$ algebra into the  $GL(1) \times SO_{n,n})$ subalgebra.  In this decomposition the simple roots of $E_{n+1}$ may be written
$$
\vec{\alpha}_{i}=\left( 0 , \tilde{\alpha}_{i} \right), \quad i=1,...,n-1, 
\quad 
\vec{\alpha}_{n}=\left( x , -\tilde{\lambda}_{n-1} \right), \quad 
\quad 
\vec{\alpha}_{n+1}=\left( 0,  \tilde{\alpha}_{n}   \right), 
\eqno(B.39)
$$
where the tilde denotes $SO(n,n)$ simple roots and fundamental weights.  The variable $x$ is fixed by the condition on the length of the simple roots, $\vec{\alpha}_{n+1}^{2}=2=x^{2}+\tilde{\lambda}_{n-1}^{2}$, this leads to 
$$
x^{2}={8-n \over 4}.
\eqno(B.40)
$$
The corresponding fundamental weights are
$$
\vec{\Lambda}^{i}=\left({\tilde{\lambda}_{i}.\tilde{\lambda}_{n-1} \over x}, \tilde{\lambda}_{i} \right), \ \ \ i=1,...,n-1, 
\quad 
\vec{\Lambda}^{n}=\left( {1 \over x}, \tilde{0} \right),  
\quad 
\vec{\Lambda}^{n+1}=\left({\tilde{\lambda}_{n}.\tilde{\lambda}_{n-1} \over x}, \tilde{\lambda}_{n} \right).
\eqno(B.41)
$$
where the tilde denotes $SO(n,n)$ simple roots and fundamental weights. 

In taking the $ g_{d} \rightarrow 0 $ limit, which is equivalent to $\varphi_{n} \rightarrow \infty$, we must fix the quantities 

$${\tilde{\varphi}}_{i}=
 {\dot \varphi}_i - \dot{\varphi}_{n} (\underline{A}^{-1})^{i \ n-1}+ \dot{\varphi}_{n+1} \underline{\alpha}_{n}, 
\eqno(B.42)
$$
for $i=1,...,n-1$ and
$$
{\tilde{\varphi}}_{n+1}= \dot{\varphi}_{n+1} - \dot{\varphi}_{n} (\underline{A}^{-1})^{i \ n} 
\eqno(B.43)
$$
where $(\underline{A}^{-1})$ is the inverse Cartan matrix of $SO(n,n)$, to preserve the $SO(n,n)$ symmetry. 

\par
In the perturbative limit $g_{d}= e^{-\left({8-n \over 4}\right) \dot{\varphi}_{n} } \rightarrow 0$ the Laplacian $\Delta$ becomes
$$
\Delta= { 1 \over 2x^{2} } { \partial \over \partial \dot{\varphi}_{n} } { \partial \over \partial \dot{\varphi}_{n} } - { (3n^{2} -n - 4) \over 2(8-n)   } { \partial \over \partial \dot{\varphi}_{n} } + \Delta_{SO(n,n)}
$$
$$
= { 4 \over 2(8-n) } { \partial \over \partial \dot{\varphi}_{n} } { \partial \over \partial \dot{\varphi}_{n} } - { (3n^{2} -n - 4) \over 2(8-n)   } { \partial \over \partial \dot{\varphi}_{n} } + \Delta_{SO(n,n)}
\eqno(B.44)
$$
where we have used $x^{2}= {8-n \over 4}$

\medskip
\noindent
{\bf {Acknowledgment}}
\medskip 
We wish to thank the
SFTC for support from the consolidated grant number ST/J002798/1.

\medskip
\noindent
{\bf References}
\medskip

\item{[1]}
I.~C.~G.~Campbell and P.~C.~West,
{\it N=2 D=10 Nonchiral Supergravity and Its Spontaneous
Compactification},
Nucl.\ Phys. B {\bf 243} (1984) 112.

\item{[2]}
F.~Giani and M.~Pernici,
{\it N=2 Supergravity in Ten-Dimensions},
Phys.\ Rev. D {\bf 30} (1984) 325.

\item{[3]}
M.~Huq and M.~A.~Namazie,
{\it Kaluza-Klein Supergravity in Ten-Dimensions},
Class.\ Quant.\ Grav.\ {\bf 2}, 293 (1985)
[Erratum-ibid.\ {\bf 2}, 597 (1985)].
\medskip
\item{[4]}
J.~H.~Schwarz and P.~C.~West,
{\it Symmetries and Transformations of Chiral N=2 D=10 Supergravity},
Phys.\ Lett. B {\bf 126}, 301 (1983).

\item{[5]}
P.~S.~Howe and P.~C.~West,
{\it The Complete N=2, D=10 Supergravity},
Nucl.\ Phys. B {\bf 238}, 181 (1984).

\item{[6]}
J.~H.~Schwarz,
{\it Covariant Field Equations of Chiral N=2 D=10 Supergravity},
Nucl.\ Phys. B {\bf 226}, 269 (1983).

\item{[7]}
M.~B.~Green and M.~Gutperle,
{\it Effects of D instantons},
Nucl.\ Phys. B {\bf 498}, 195 (1997)
[arXiv:hep-th/9701093].

\item{[8]}
M.~B.~Green, M.~Gutperle and P.~Vanhove,
{\it One loop in eleven-dimensions},
Phys.\ Lett. B {\bf 409}, 177 (1997)
[arXiv:hep-th/9706175].

\item{[9]}
M.~B.~Green and S.~Sethi,
{\it Supersymmetry constraints on type IIB supergravity},
Phys.\ Rev. D {\bf 59}, 046006 (1999)
[arXiv:hep-th/9808061].

\item{[10]}
M.~B.~Green, H.~h.~Kwon and P.~Vanhove,
{\it Two loops in eleven-dimensions},
Phys.\ Rev. D {\bf 61}, 104010 (2000)
[arXiv:hep-th/9910055].

\item{[11]}
M.~B.~Green and P.~Vanhove,
{\it Duality and higher derivative terms in M theory},
JHEP {\bf 0601}, 093 (2006)
[arXiv:hep-th/0510027].

\item{[12]}
M.~B.~Green, J.~G.~Russo and P.~Vanhove,
{\it Modular properties of two-loop maximal supergravity and
connections with
string theory},
JHEP {\bf 0807}, 126 (2008)
[arXiv:0807.0389 [hep-th]].

\item{[13]}
J.~G.~Russo,
{\it Construction of SL(2,Z) invariant amplitudes in type IIB
superstring
theory},
Nucl.\ Phys. B {\bf 535}, 116 (1998)
[arXiv:hep-th/9802090].

\item{[14]}
A.~Basu,
{\it The D**10 R**4 term in type IIB string theory},
Phys.\ Lett. B {\bf 648}, 378 (2007)
[arXiv:hep-th/0610335].

\item{[15]}
E.~Kiritsis and B.~Pioline,
{\it On R**4 threshold corrections in IIb string theory and (p, q)
string
instantons},
Nucl.\ Phys. B {\bf 508}, 509 (1997)
[arXiv:hep-th/9707018].
  
\item{[16]}
N.~A.~Obers and B.~Pioline,
{\it Eisenstein series and string thresholds},
Commun.\ Math.\ Phys. {\bf 209}, 275 (2000)
[arXiv:hep-th/9903113]; N.~A.~Obers and B.~Pioline,
{\it Eisenstein series in string theory},
Class.\ Quant.\ Grav. {\bf 17}, 1215 (2000)
[arXiv:hep-th/9910115].

\item{[17]}
  A.~Basu,
  {\it The $D^4 R^4$ term in type IIB string theory on $T^2$ and U-duality},
  Phys.\ Rev.\  D {\bf 77}, 106003 (2008)
  [arXiv:0708.2950 [hep-th]].
  
\item{[18]}
  A.~Basu,
  {\it The $D^6 R^4$ term in type IIB string theory on $T^2$ and U-duality},
  Phys.\ Rev.\  D {\bf 77}, 106004 (2008)
  [arXiv:0712.1252 [hep-th]].

\item{[19]}
N.~Lambert and P.~West,
{\it Perturbation Theory From Automorphic Forms},
JHEP {\bf 1005}, 098 (2010)
[arXiv:1001.3284 [hep-th]].

\item{[20]}
M.~B.~Green, J.~G.~Russo and P.~Vanhove,
{\it Automorphic properties of low energy string amplitudes in
various
dimensions},
Phys.\ Rev. D {\bf 81}, 086008 (2010)
[arXiv:1001.2535 [hep-th]].

\item{[21]}
M.~B.~Green, J.~G.~Russo and P.~Vanhove,
{\it String theory dualities and supergravity divergences},
JHEP {\bf 1006}, 075 (2010)
[arXiv:1002.3805 [hep-th]].

\item{[22]}
M.~B.~Green, S.~D.~Miller, J.~G.~Russo and P.~Vanhove,
{\it Eisenstein series for higher-rank groups and string theory
amplitudes},
arXiv:1004.0163 [hep-th].

\item{[23]}
B.~Pioline,
{\it R**4 couplings and automorphic unipotent representations},
JHEP {\bf 1003}, 116 (2010)
[arXiv:1001.3647 [hep-th]].

\item{[24]} 
  A.~Basu,
  {\it The structure of the $R^{8}$ term in type IIB string theory},
  arXiv:1306.2501 [hep-th].
  
\item{[25]} 
  A.~Basu and S.~Sethi,
  {\it Recursion Relations from Space-time Supersymmetry},
  JHEP {\bf 0809}, 081 (2008)
  [arXiv:0808.1250 [hep-th]].
  
\item{[26]} 
  A.~Basu,
  {\it Constraining gravitational interactions in the M theory effective action},
  arXiv:1308.2564 [hep-th].
  
  \item{[27]}
F.~Gubay, N.~Lambert, P.~West,
{\it Constraints on Automorphic Forms of Higher Derivative Terms
from Compactification},
JHEP {\bf 1008}, 028 (2010).
[arXiv:1002.1068 [hep-th]]

\item{[28]} 
  F.~Gubay and P.~West,
  {\it Higher derivative type II string effective actions, automorphic forms and $E_{11}$},
  JHEP {\bf 1204}, 012 (2012)
  [arXiv:1111.0464 [hep-th]].

\item{[29]} 
  F.~Gubay and P.~West,
  {\it Parameters, limits and higher derivative type II string corrections},
  JHEP {\bf 1211}, 027 (2012)
  [arXiv:1204.1403 [hep-th]].

\item{[30]}
  P.~C.~West,
  {\it Introduction to strings and branes},
	{\it  Cambridge, UK: Cambridge University Press (2012) 672p}.

\item{[31]}
	M.~B.~Green, J.~G.~Russo and P.~Vanhove,
	{\it Non-renormalisation conditions in type II string theory and
	maximal supergravity},
	JHEP {\bf 0702}, 099 (2007) [arXiv:hep-th/0610299].

\item{[32]}
  M.~R.~Gaberdiel, D.~I.~Olive, P.~C.~West,
  {\it A Class of Lorentzian Kac-Moody algebras},
  Nucl.\ Phys.\  {\bf B645}, 403-437 (2002).
  [hep-th/0205068].

\end